%% file: ms.tex
\documentclass[sigconf]{acmart}

\pdfoutput=1

\usepackage[utf8]{inputenc}
\usepackage{xcolor}

\DeclareGraphicsExtensions{.pdf,.png,.jpg}

\input{macros}

\begin{document}

\title{Onions in the Crosshairs}
\subtitle{When The Man really \emph{is} out to get you}

\acmConference{}{}{}

 \author{Aaron D. Jaggard}
 \orcid{0000-0003-0628-4553}
\affiliation{U.S. Naval Research Laboratory}
 \email{aaron.jaggard@nrl.navy.mil}

 \author{Paul Syverson}
 \orcid{}
 \affiliation{U.S. Naval Research Laboratory}
 \email{paul.syverson@nrl.navy.mil}

\setcopyright{none}
\settopmatter{printacmref=false}
\acmDOI{}
\acmYear{}
\acmPrice{}
\acmISBN{}

\begin{abstract}
We introduce and investigate \emph{targeting adversaries} who selectively attack users of Tor or other secure-communication networks. We argue that attacks by such adversaries are more realistic and more significant threats to those most relying on Tor's protection than are attacks in prior analyses of Tor security. Previous research and Tor design decisions have focused on protecting against adversaries who are equally interested in any user of the network. Our adversaries selectively target users---e.g., those who visit a particular website or chat on a particular private channel---and essentially disregard Tor users other than these.  We present a model of such adversaries and investigate three example cases where particular users might be targeted: a cabal conducting meetings using MTor, a published Tor multicast protocol; a cabal meeting on a private IRC channel; and users visiting a particular .onion website. In general for our adversaries, compromise is much faster and provides more feedback and possibilities for adaptation than do attacks examined in prior work. We also discuss selection of websites for targeting of their users based on the distribution across users of site activity. We describe adversaries both attempting to learn the size of a cabal meeting online or of a set of sufficiently active visitors to a targeted site and attempting to identify guards of each targeted user. We compare the threat of targeting adversaries versus previously considered adversaries, and we briefly sketch possible countermeasures for resisting targeting adversaries.
\end{abstract}

\maketitle

\input{intro}

\input{shortadv}
\input{mtor}
\input{irc}
\input{profile}
\input{discussion}
\input{conc}

\section*{Acknowledgments}
We thank Giulia Fanti, Nick Hopper, Rob Jansen, George Kadianakis, Matt Traudt, and Ryan Wails for their comments on earlier drafts of this paper.

\appendix
\input{model}
\input{adv}
\input{irccomp}
\input{dmle}

\end{document}

%% file: macros.tex
\newcommand{\Ie}{{I.e.}}
\newcommand{\eg}{{e.g.}}
\newcommand{\ie}{{i.e.}}
\newcommand{\etc}{{etc.}}
\newcommand{\etal}{{et al.}}

\renewcommand{\paragraph}[1]{\noindent{\textbf{#1:}}}

\newcommand{\princ}{\ensuremath{\mathcal{P}}}
\newcommand{\identities}{\ensuremath{\mathcal{I}}}
\newcommand{\clients}{\ensuremath{\mathcal{CL}}}
\newcommand{\relays}{\ensuremath{\mathcal{R}}}
\newcommand{\guards}{\ensuremath{\mathcal{G}}}
\newcommand{\guardset}{\ensuremath{G}}

\newcommand{\middles}{\ensuremath{\mathcal{M}}}
\newcommand{\exits}{\ensuremath{\mathcal{E}}}
\newcommand{\circs}{\ensuremath{\mathcal{CIR}}}
\newcommand{\dests}{\ensuremath{\mathcal{D}}}
\newcommand{\links}{\ensuremath{\mathcal{L}}}
\newcommand{\gdist}{\ensuremath{\gamma}}
\newcommand{\ddist}{\ensuremath{\delta}}
\newcommand{\tr}{\ensuremath{\mathsf{tr}}}
\newcommand{\ci}{\ensuremath{\mathsf{ci}}}
\newcommand{\cl}{\ensuremath{\mathsf{cl}}}
\newcommand{\targets}{\ensuremath{\mathcal{T}}}
\newcommand{\target}{\ensuremath{T}}
\newcommand{\pred}{\ensuremath{\mathsf{P}}}
\newcommand{\profile}{\ensuremath{\mathcal{P}}}

\newcommand{\act}{\ensuremath{A}}
\newcommand{\aux}{\ensuremath{\mathsf{Aux}}}
\newcommand{\card}[1]{\ensuremath{\vert{#1}\vert}}

\theoremstyle{remark}
\newtheorem{remark}[theorem]{Remark}

%% file: intro.tex
\section{Introduction}\label{sec:intro}

Tor is a network for traffic security of Internet
communications~\cite{tor-design} with millions of
users~\cite{tormetrics}. Most Tor users are unlikely to be of
specific interest to an adversary; they are primarily protected by Tor
against opportunistic local eavesdroppers and local censors or against
hostile destinations. Deanonymizing adversaries are generally modeled
as attempting to attack as many users as possible rather than targeting
particular users or groups.

For many Tor users this is perhaps appropriate, but Tor is explicitly
intended to protect human rights workers, law enforcement, military,
journalists, and others~\cite{who-uses-tor-url} who may face large,
well-financed, and determined adversaries. More to the point, some of
these adversaries adversaries will hoover up whatever they can, but
they may also be more interested in specific individuals or groups of
Tor users, possibly based on offline or out-of-band reasons. An
adversary whose interest is directed primarily or more fervently at
particular users may employ different strategies. And if Tor's design
decisions are motivated by analyses of what hoovering adversaries can
do, those most in need of Tor's protections may be the least well
served.

We regard the adversaries in this paper as the next step in an
ongoing evolution of most appropriate and important onion routing
adversaries, away from abstracting reality till it matches models
and towards better matching models to reality. 
Our focus in this work is on \emph{targeting adversaries}.  These need
not differ at all from previously studied adversaries in terms of
their capabilities or resource endowment, though they might. They
differ primarily in their goals and strategies. We will set out
various types of targeting adversaries presently; however, we mention
an example here to give the basic idea.  A targeting (or ``selective'') adversary, Sam,
who has compromised a particular user of interest, Alice, and observed
her connecting to Bob, an interesting and unusual .onion website
(essentially websites reachable only over Tor) may wish to target
other users of that site. Sam might be particularly interested to
learn which are the most active site users or how popular the site is
in general.

\noindent {\bf Background:} We sketch here a basic background on Tor
to provide context for this work.  For further
descriptions, see the Tor design paper~\cite{tor-design}, or related
documentation at the Tor website~\cite{torproject}.  Tor clients
randomly select sequences of three out of roughly 10,000 relays~\cite{tor-network-size}
forming the current Tor network, and create a cryptographic circuit
through these to connect to Internet services. Since only the first
relay in the circuit sees the IP address of the client and only the
last (exit) relay sees the IP address of the destination, this
technique separates identification from routing. In response to a
vulnerability analysis~\cite{hs-attack06}, Tor began having clients
each choose a small set of \emph{entry guards} from which to
persistently choose the first relay in any circuit. The appropriate
size and rotation criteria for the set of guards is the focus of
ongoing work, including that presented below. Tor currently has
roughly two million users connecting in this
way~\cite{tormetrics}. For simplicity of this first analysis of
targeting adversaries, we ignore the 150,000 users currently connecting
to Tor via bridges, a mechanism to access Tor in environments that
censor users connecting to the Tor network at all. Nonetheless, much
of our analysis will apply there as well.

Tor also facilitates onion services, which are services at Internet
sites on the .onion top-level domain that is reserved by IETF standard~\cite{ietf-onion-tld-rfc}. Onionsites create circuits into
the Tor network to \emph{Introduction Points} at which they are then
reachable. Thus, though not part of the Tor network itself, onionsites
are only reachable via Tor.  Connections to them, including for address lookup, do not exit the Tor network described above.

\noindent{\bf Result Highlights:} MTor is a published protocol for
multicast over Tor~\cite{mtor-popets16}. Its security goals are to
``prevent an adversary from (1) discerning the sender of the message
and (2) enumerating group members.'' We show in Sec.~\ref{sec:mtor}
that a targeting
adversary with capabilities within those assumed by MTor's authors can
enumerate group members and can identify the guard relay of a message
sender. 

In Sec.~\ref{sec:irc}, we describe how a targeting
adversary will, within a few weeks of attack initiation, have a high
probability of locating the leader of a cabal that meets several times
per day on a private IRC channel. In contrast, to achieve the same
expectation of cabal-leader location by an adversary with roughly the
same resources using previous strategies would require several
months~\cite{jwjss13ccs}. We also show that targeting adversaries
receive feedback on intermediate goals such as cabal size and
activity, allowing them to decide, adapt, or refocus subsequent
attacks at a similarly faster rate than previous adversaries. 
Those results are discussed in Sec.~\ref{sec:discussion}, where
we will also briefly describe possible counters to the kinds of targeted
attacks we introduce in this paper.  

Tor has recently taken steps to make it difficult for adversaries
to predict the onionsites for which relays they own would function as
directory~\cite{prop250} or recognize which onionsite is being
requested when receiving a directory request~\cite{prop224}. This was
in part to counter published attacks allowing an adversary to monitor
interest in onionsites by monitoring the rate of directory
requests~\cite{Biryukov-2013}. Using attacks similar to those that we
describe against MTor and IRC cabals, we show in
Sec.~\ref{sec:profile} that a moderately resourced adversary can
assess not just the level of site activity but the distribution of
client interaction with a targeted onionsite and will identify the
guards of more active clients, potentially for additional targeted
attacks. 

After noting some relevant prior work next, we present in
Sec.~\ref{sec:shortadv} a brief description of the targeting
adversaries used in the worked examples just mentioned, along with the
general strategy they all follow. A more general and abstract
description of various types of targeting adversaries, their goals,
and their properties is reserved for
Apps.~\ref{app:model}~and~\ref{app:adv}.

\noindent{\bf Related Work:} We will primarily discuss related work at
points in the text where that work is relevant.  We here briefly
mention a few highlights of prior general work on Tor (or more generally,
onion routing) adversary models and security analysis.

Analysis of onion routing security has generally focused on end-to-end
correlation. To be practical, onion routing networks are generally
low-latency. Thus, an adversary able to observe both ends of a
connection can correlate patterns of communication and correlate
connection source and destination with little error regardless of what
happens between the ends. Given the fraction $f$ of Tor relays that
are compromised or observed, this provides roughly $f^2$ probability
of any one onion-routing circuit being
compromised~\cite{strl01correlation}.  Various end-to-end
correlating adversaries and
this metric of security against them are the basis for the bulk of Tor
security research and design.  Hintz, however, was the first to
observe that if an adversary can recognize a destination from the
usual pattern of traffic that results from connecting to it, then it
is sufficient to observe the client end to recognize its destination
in a \emph{fingerprinting} attack~\cite{hintz:pet2002}. Such
recognition is a central feature of attacks for all three of our
examples.

Feamster and Dingledine were the first to examine an adversary
occupying the network links between relays rather than at the relays
themselves~\cite{feamster:wpes2004}. Vulnerability to link adversaries
is significant enough that any useful Tor security design must take
them into account. Nonetheless, we will show that a targeting relay
adversary is sufficient to carry out effective attacks.

Prior to the last half decade, research has primarily looked at the risk of
correlation at a network snapshot. Johnson et al.\ 
considered security of using the Tor network over time, examining such
questions as the time until a user with a given type of behavior is likely to
experience its first correlated connection, and given such behavior,
the fraction of connections that will be so compromised over a period
of use~\cite{jwjss13ccs}. Since they consider IRC use
as one of their classes of behavior, we will compare the attacks
we devise on an IRC cabal to those they examined.

Not all work prior to Johnson et al.\ ignored observation over time.
Predecessor and intersection attacks examine repeated connections or
message transmissions to see who is the only one or the most common
one who could have been sending when a given destination is
receiving. Crowds was a published system design for anonymous web
browsing that was created with these attacks in
mind~\cite{crowds}. Wright et al.\ analyzed these attacks for many
traffic security systems including pre-Tor onion
routing~\cite{wright03}.  Most intersection attacks and analyses
looked for \emph{any} association of a sender and receiver. As such they were
not targeted. However, the first such attacks conducted on a deployed,
publicly-used system were used to show that an adversary could
repeatedly connect to and thereby find the IP address of a specific
hidden onion service. These were a basis for introducing guard relays
to Tor~\cite{hs-attack06}. Nonetheless, end-to-end correlation still
remains the primary concern of most Tor analyses and defenses.

%% file: shortadv.tex
\section{Targeting Adversaries}\label{sec:shortadv}

We expect selective targeting Tor adversary description and analysis
to be amenable to rigorous formal reasoning. We also anticipate future
analyses of other targeting adversaries than in our examples, \eg, an
adversary that attempts for a given targeted user to build a profile
of that user's selected destinations and activity at them over a given
period.  To this end, we set out in Apps.~\ref{app:model}
and~\ref{app:adv} an abstract model of both system elements and
actions, as well as different categories of targeting adversaries and
their goals. Here we simply sketch the basic adversary properties and
strategy that should apply to all of our worked examples.  In the
first two of these examples, the adversary is interested in a cabal of
users communicating through Tor with either a multicast system or IRC;
in our third main example, he is interested not in a cabal per se, but
in the set of users who frequently visit a targeted onionsite.

The general approach that all of our examples follow is to have an
adversary that initially deploys all of its relay capacity at middle
relays in the network. We assume that communication within a targeted
cabal or with a targeted onionsite is recognizable by its traffic patterns
at middle relays.  The basis of that assumption varies
with example and is stated in each section. The initial strategy of
the adversary is then to attempt to find a guard for each of the
targeted clients by observing which guards transmit or receive
recognizable target traffic. This may be a final strategy if the
adversary's only goal is to learn the size of a cabal and/or monitor
the distribution of its activity.  But, it may be
just a stepping stone, \eg, to inform the decision whether to attempt
to ``bridge'' guards---\ie, to transition beyond knowing that a guard is
being used by one or more clients of interest to identifying client IP addresses.
The adversary may be selective in this decision as well; rather than targeting
all cabal members it might, \eg, attempt to bridge only for those that 
send the most or prompt the most responses
when they send.  Note that ``bridging'' typically implies getting across
a network obstacle, rather than compromising it directly~\cite{ds08pets}.
In our setting this could be done via the guard ISP, compromised ASes
between a guard and client, etc. For convenience, we will subsume
under ``bridging'' both bridging in that sense and compromise of a guard
by physical access to its hardware, threatening its operator, exploiting configuration errors, using zero-day exploits on its software,
etc.

Another adversary goal is to
assign confidence to its assessment of cabal size. This can involve
evaluation of confidence in the correctness of the fraction of cabal
users who have had a guard identified. The adversary will also need
to evaluate the expected degree of guard collision amongst cabal
members.

%% file: mtor.tex
\section{Example: Multicast Cabals}
\label{sec:mtor}

MTor is a design for multicast group communication over Tor recently introduced by Lin~\etal~\cite{mtor-popets16}.

\subsection{MTor overview}

Each client that joins a
given MTor multicast group creates a circuit to a Tor relay that serves as
the group's current multicast root (MR).  We do not describe MTor's
selection or rotation of MR and skip many other details as well.
Communication for the group travels up this circuit and then
propagates down from any node with untraversed subtrees to the group
members at the leaves. This creates the overhead savings in Tor
circuit construction of a tree (at largest a star topology) versus
pairwise connections to all group members. As we will see, for moderately
sized cabals on the current Tor network, a star topology or a tree
that is almost a star is reasonable to expect.

Since the MR is a Tor relay, MTor's design also incurs the performance
and overhead advantages of not having traffic exit the
network. Limited network exit capacity is often a dominating factor
for Tor performance and is one of the motivations for Facebook's
offering an onionsite rather than merely encouraging connections to their
registered domain via Tor~\cite{7686-and-all}. And, MTor circuits
are client--guard--middle--MR vs.\ client--guard--middle--exit-server for unicast,
thus saving one hop of path overhead for each group member.

MTor also modifies normal Tor behavior for the potential performance
gain from message deduplication that is typical of multicast.  Tor normally
creates cryptographic circuits by tunneling a Diffie-Hellman protocol
to establish a session key known only to the client and to the next
onion router in the circuit being built. MTor uses group keys so that
if a client attempts to build a circuit through a relay that is
already part of the same multicast tree, the relay will recognize the
session group identifier (GID) sent by the client and join the new circuit
to the existing group circuit rather than continue building as per the
client request. To further manage tree size and improve the advantages
of multicast, MTor also allows the restriction of middle relay
selection for MTor communication to a designated subset of relays
and/or to relays having a minimum bandwidth.

\subsection{MTor adversary}
A targeted and compromised Alice belonging to a cabal that meets only
via MTor reveals to the adversary all cabal communications, as well
as long-term group keys and identifiers.  A targeted but uncompromised Alice with a compromised guard connecting
to an MTor cabal could make the cabal a target by association. (When
the targeted group is open, the adversary can join it too.)

\subsubsection{Adversary goals}
Lin~\etal\ consider adversary goals of looking at all pairs of users
and trying to link each pair as part of a multicast group (by their
guards seeing the same GID) and of identifying a user as participating
in a multicast group (by a guard seeing the GID---experiments consider
only a single multicast group)~\cite{mtor-popets16}. 
While these may be useful for some
purposes, our targeting adversary has goals of identifying all members
of a multicast group of interest, estimating the cabal's size, or
identifying MTor groups to which a targeted user might
belong. Indeed, a targeted user communicating over MTor is a natural
subject of all the adversary goals identified in
App.~\ref{app:adversary-goals}.

\subsubsection{Adversary endowment and capabilities}
For simplicity, and like Lin~\etal, we will consider only a relay
adversary.  On the other hand, it will be useful for our adversary to
compromise relays other than guards. An adversary that owns middle
relays can both estimate the cabal size and identify guards to target for compromise or bridging
so as to identify the clients behind them.
Even the MR can estimate cabal size. 

The guard of an MTor group member can see all session GIDs for the
group, and may then wish to identify, \eg, others in that group. We
will assume, however, that traffic patterns for any cabal member in a
multicast session will be adequately linkable so that the GID will not
be needed for this adversary to associate other clients with the
cabal. (This assumption is also made by MTor's authors.)
In general, we consider an adversary capable of active
attacks, including disrupting group communications or generating
its own group traffic if a member. For simplicity, however, our
initial analysis assumes a passive adversary. Since MTor sessions are
always identifiable by a participating adversary relay and our
analysis will parametrize over the number of sessions, this is not as
significant a limitation on the adversary as is usually the case for
Tor communications.

\subsection{MTor cabal analysis}

Note that while the GID is not that significant to a targeting
adversary who can observe traffic patterns, the multicast tree
structure is. To illustrate with an unrealistic example, if the
middle-relay set were restricted to a singleton, then in sessions
where the adversary has compromised this relay he has thereby
identified all the guards used by any cabal members in that
session.  Lin et al.\ do not give criteria for middle-relay-set
restriction. If gameable an adversary might be able to improve its
expected inclusion in this set disproportionate to its actual relay
and bandwidth resources.  On the other hand, a restriction of
middle-relay-set size can obscure cabal size estimates by a
an adversarial MR\@.

\subsubsection{Learning a guard of every cabal member}\label{ssec:mtor-guard}

For our initial analysis, we consider an adversary who controls a
fraction $B$ of the middle-relay bandwidth and seeks to identify at least
one guard of each member of a cabal.  (This might be used to
adaptively target guards for future attack or as part of an estimation
of cabal size. If the adversary has a joined or compromised cabal member,
he can associate a guard with every cabal member who ever sends. ) 
We also consider the effects of the number $c$ of
cabal members and the number $m$ of cabal multicast sessions
observed.  If the instance of MTor restricts the set of usable middle
relays to obtain the associated deduplication benefit, we take $B$ to be the
fraction of MTor-available middle-relay bandwidth that is controlled by the adversary.  Here, we also consider a probability $T$ that an adversary might allow for his failure.

We assume for simplicity a static network that does not change for the
period of our analysis.  In each multicast session, a new random MR is
chosen and the circuits constituting the multicast tree are also
reformed.  We also assume that
all members of the cabal participate in all multicast sessions
(meetings) and that cabal composition does not change.

If a cabal member constructs a circuit that uses a middle relay controlled by the adversary, then the adversary learns the client's guard for that circuit.  We take this as the only way that the adversary learns guards.

\begin{figure}
\includegraphics[width=0.5\textwidth]{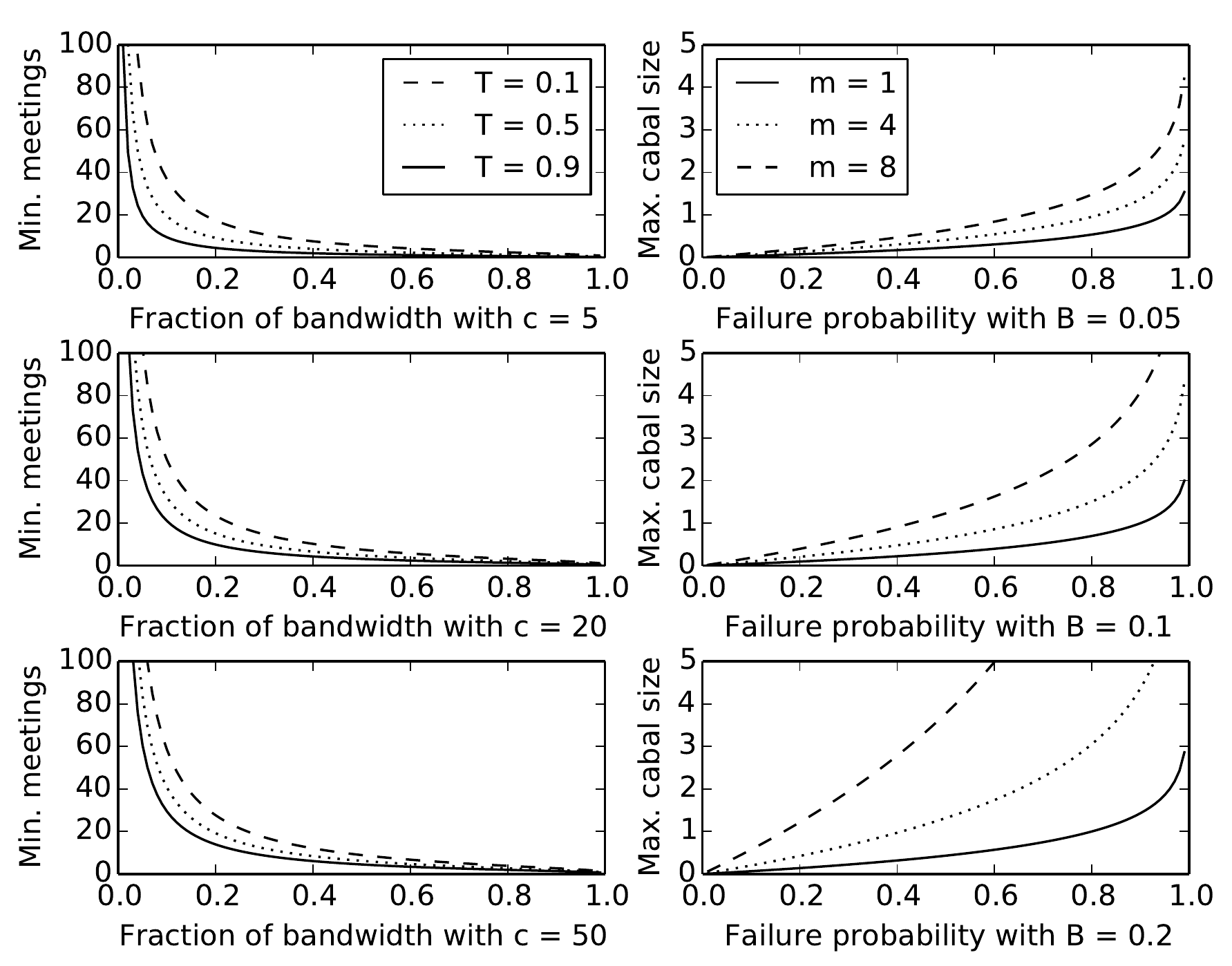}
\caption{Left column: The minimum number of meetings $m$, as a function of middle-relay bandwidth $B$ controlled by the adversary, required for the adversary to learn at least one guard of each cabal member for cabal sizes $c=5$, $20$, and $50$ (top to bottom).  Different curves in each subplot correspond to different failure probabilities for the adversary as indicated.  Right column: The maximum cabal size $c$, as a function of the adversary's failure threshold $T$, required for the adversary to learn at least one guard of each cabal member when the adversary controls $B=0.05$, $0.1$, and $0.2$ of the middle-relay bandwidth (top to bottom).  Different curves in each subplot correspond to different numbers of cabal meetings as indicated.}
\label{fig:mtor-basic-analytic}
\end{figure}

The probability that a given cabal member never uses a compromised
middle relay in any of $m$ sessions is $(1-B)^m$.  Given the
simplifying assumption that such compromise is independent for all
clients, the
probability that a guard of every one of the $c$ cabal members is identified
at least once over the course of $m$ meetings is thus $(1-(1-B)^m)^c$.
We can then gauge adversary success by bounding the probability
that the adversary fails to carry out this compromise, giving us
\begin{equation}
1 - \left(1 - \left(1 - B\right)^m\right)^c < T.
\label{eq:mtor-BmcT}
\end{equation}
We now explore the parts of the $(c,m,B,T)$ space that satisfy this inequality.
If the number of meetings
\begin{equation}
m > \log_{1-B} \left[1 - \left(1-T\right)^{1/c}\right],
\label{eq:mtor-m}
\end{equation}
where $B\in(0,1)$, then, with probability at least $1-T$, the
adversary learns at least one guard of each of the $c$ cabal
members.  The left column of Fig.~\ref{fig:mtor-basic-analytic} plots the
right-hand side of (\ref{eq:mtor-m}) as a function of $B$ for $c=5$, $20$,
and $50$ (top to bottom subplots) and $T = 0.1$, $0.5$, and $0.9$ as
indicated on the curves within each subplot. Thus, for cabal sizes up
to $50$ an adversary holding $10\%$ of middle-relay bandwidth 
has a $90\%$ chance of having identified a guard for each
cabal member after no more than $60$ meetings. On the other hand, after
$10$ meetings, even for $c=5$ the adversary must hold a more ambitious
middle-relay bandwidth of nearly $40\%$. Note that the more 
relays the adversary introduces the more gradually he must introduce
them and the more the relays should be in multiple locations if he is
to avoid suspicion.

Again using~(\ref{eq:mtor-BmcT}), if the number of cabal members
\begin{equation}
c < \log_{1-\left(1 - B\right)^m} \left[1 - T\right],
\label{eq:mtor-c}
\end{equation}
where $B\in (0,1)$ and $m$ is a positive integer, then the adversary will learn at least one guard of each of the $c$ cabal members.  The right column of Fig.~\ref{fig:mtor-basic-analytic} plots the right-hand-side of~(\ref{eq:mtor-c}) as a function of $T$ for $B=0.05$, $0.1$, and $0.2$ (top to bottom subplots) and $m=1$, $4$, and $8$ as indicated on the curves within each subplot.

\begin{figure}
\includegraphics[width=0.45\textwidth]{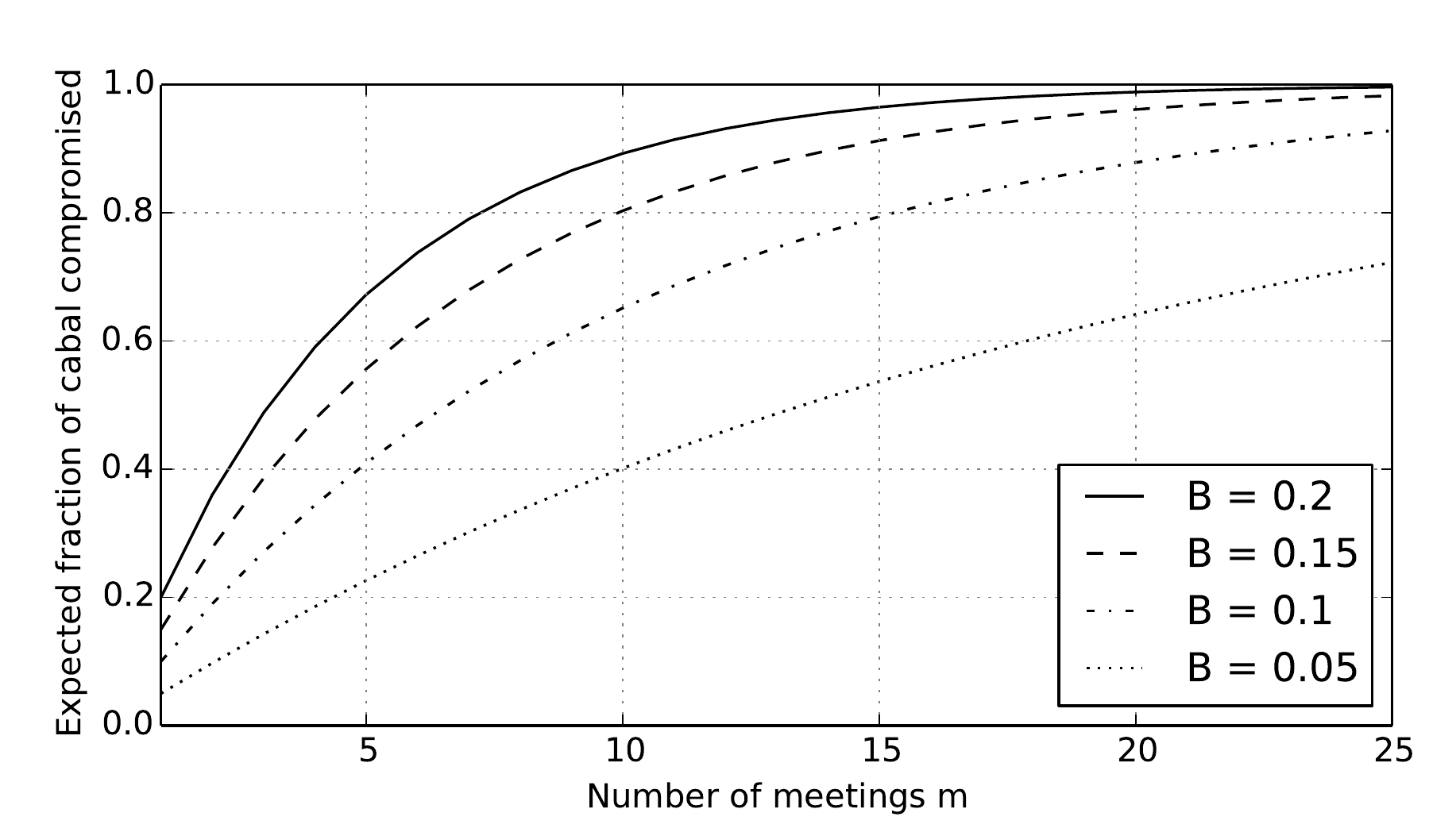}
\caption{The expected fraction of the cabal members for which the adversary identifies at least one guard as a function of the number of meetings $m$.  Different curves show different fractions $B$ of middle-relay bandwidth controlled by the adversary.}
\label{fig:mtor-comp-fraction}
\end{figure}

From the cabal's perspective, it should keep its set of members
minimal with respect to the qualities needed to accomplish its goals.  However, the
cabal might reasonably ask how many meetings it should hold and what
the effects of additional meetings are on its security.  For a given
cabal size $c$ and a subset of $i$ cabal members, the probability that
a guard of exactly those $i$ members is exposed after $m$ meetings is
$\left(1-(1-B)^m\right)^i \left(1-B\right)^{m(c-i)}$.
Considering the different ways to choose the subset of cabal members, the expected number of cabal members with an identified guard after $m$ meetings is
$\sum_{i=0}^c i \binom{c}{i} \left(1-(1-B)^m\right)^i \left(1-B\right)^{m(c-i)}$.  
Applying known results about the mean of the binomial distribution, the expected number of cabal members with no guard identified is $c(1-B)^m$.  The fraction
of the cabal that has identified guards is thus independent of $c$ and
equal to $1-(1-B)^m$.  Figure~\ref{fig:mtor-comp-fraction} shows the
expected fraction of cabal members with identified guards after $m$
meetings for different fractions $B$ of middle-relay bandwidth
controlled by the adversary.
Perhaps unsurprisingly, failure probability appears to be highly responsive
to either the number of meetings observed or the fraction doing the observing.

\subsubsection{Estimating cabal size as multicast root}
\label{cabalsize}

An adversary who controls the MR can estimate cabal size from the
number of distinct GID circuits connecting to the MR\@.
The MR is selected uniformly at random from all relays
that have the fast and stable flags and meet the minimum bandwidth
parameter of the multicast group. We assume that
all adversary relays meet these requirements. Given a fixed fraction of
middle-relay adversary bandwidth, the adversary maximizes his chances
of being selected for MR if he distributes that bandwidth across as
many relays as possible, subject to inclusion requirements. He
can accomplish this with a uniform distribution amongst this maximal set
of relays, which will also simplify any
combination of calculations based on adversary's fraction of relays
with those based on adversary's fraction of bandwidth.

As we have noted, increased deduplication decreases the information
an MR gets about cabal size from the number of observed circuits.
Collisions can occur in either the selection of guard or of middle
relay; as there are far fewer guards than middles in the
current Tor network, guard selection is by far the dominating source
of collision. For the following
calculations we assume that guard bandwidth is also uniformly distributed. 

With $g$ guards in the network, for a cabal of size $c$, the expected
number of guards selected is \mbox{$g(1-{(1 - 1/g)}^c)$}. As of this
writing, the Tor network has about 2500 guards, and Tor 
selects middles from about 5000 relays. If we assume that the expected
number of guards are selected, and middles are selected uniformly,
then the expected number of middles is given by the the same formula
as above where $c$ is replaced by $h$, the expected number of guards
selected, and guards are replaced by middles, viz:
\mbox{$m(1-{(1-1/m)}^h)$}. Thus the expected number of middles
selected, hence observed by the MR, is the same as cabal size for
cabals up to $c=41$, off by less than one for cabals up to $c=58$
and within five percent for cabals up to $c=182$.

A significantly skewed guard distribution would impact
the likelihood of collisions, again diminishing information about cabal
size available to the MR, but it may also help the adversary in narrowing
the set of most useful guards to target for bridging. 
Even for a selection among 800 guards and 2000 relays, the
above calculation gives an expected number of middles the same as cabal
size up to $c=24$ and within five percent up to $c=60$.
Whether the distribution is skewed or not, an adversary owning a
cabal member and the MR can recognize all colliding circuits for
cabal members who ever send during a meeting, which can be used
to improve the cabal size estimate.

If all relays selectable as middles by
Tor meet the group criteria for serving as MR, then the probability
of an adversary being in a position to make the above observation
is governed by the fraction of middle relays he owns. Thus, a
$B=0.2$ adversary will be more likely than not to have been chosen as
MR after only four multicast sessions and will have a nearly 90\%
chance of being chosen after 10 sessions.

\subsubsection{Estimating cabal size from middle relays}

Estimates based on relays serving as middles (second hops) in MTor
sessions are also possible. Compared to results from compromising a
multicast root, these will be much less likely to have information about the
circuits of all cabal members but much more likely to
provide some information about cabal size every session. In addition,
middle relays will identify guards of cabal members with every
observed cabal connection.

\paragraph{Setting and assumptions}
Here, we look at some very basic numerical simulations of the information learned by the adversary from middle relays in the MTor usage scenario.  These make assumptions that parallel those made in our analysis above, but they allow us to study the effects of MTor deduplication more easily.  In particular, we assume a static network and that cabal members choose guards and middle relays uniformly at random from sets of 2500 and 5000 such nodes, respectively.

\paragraph{Approach}
In each of 10,000 trials, we identify some middle relays as compromised; each is compromised independently with probability $B$, and the compromise status remains unchanged throughout $m$ cabal meetings.  We then choose a set of guards (the size of which depends on the simulation) for each cabal member; these sets are unchanged throughout the $m$ meetings.

For each meeting, each guard selects a client from its set.  For each guard selected, we then select a middle relay to use to connect to the MR.  (This simulates the choice of a middle relay made by the first client using that guard in that meeting.  Any other clients who use that guard for that meeting will use the same middle relay.)  If that middle relay is compromised and the adversary has not attempted to bridge the guard before (\ie, that guard has not been used in conjunction with a compromised middle relay in this trial), then we bridge the guard with probability $p_b$.  Across all meetings in the trial, we keep lists of the guards that have been bridged and that the adversary has tried but failed to bridge.  After determining which guards are newly compromised in each meeting, we determine the clients that have been identified; this set is the union of the clients that were compromised before the current meeting and all of those that, in the current meeting, used a guard that has been successfully bridged during any meeting up to this point (including the current meeting).

As noted in Sec.~\ref{sec:shortadv}, a guard might be bridged in many ways, and we expect an adversary to try them
in parallel and/or successively. He will likely start with the least
costly, least likely to raise suspicion, and most likely to succeed,
perhaps with varied emphasis depending on setting.  We use $p_b$ to
capture the cumulative probability of success of these different
approaches.

\begin{figure}
\includegraphics[width=0.45\textwidth]{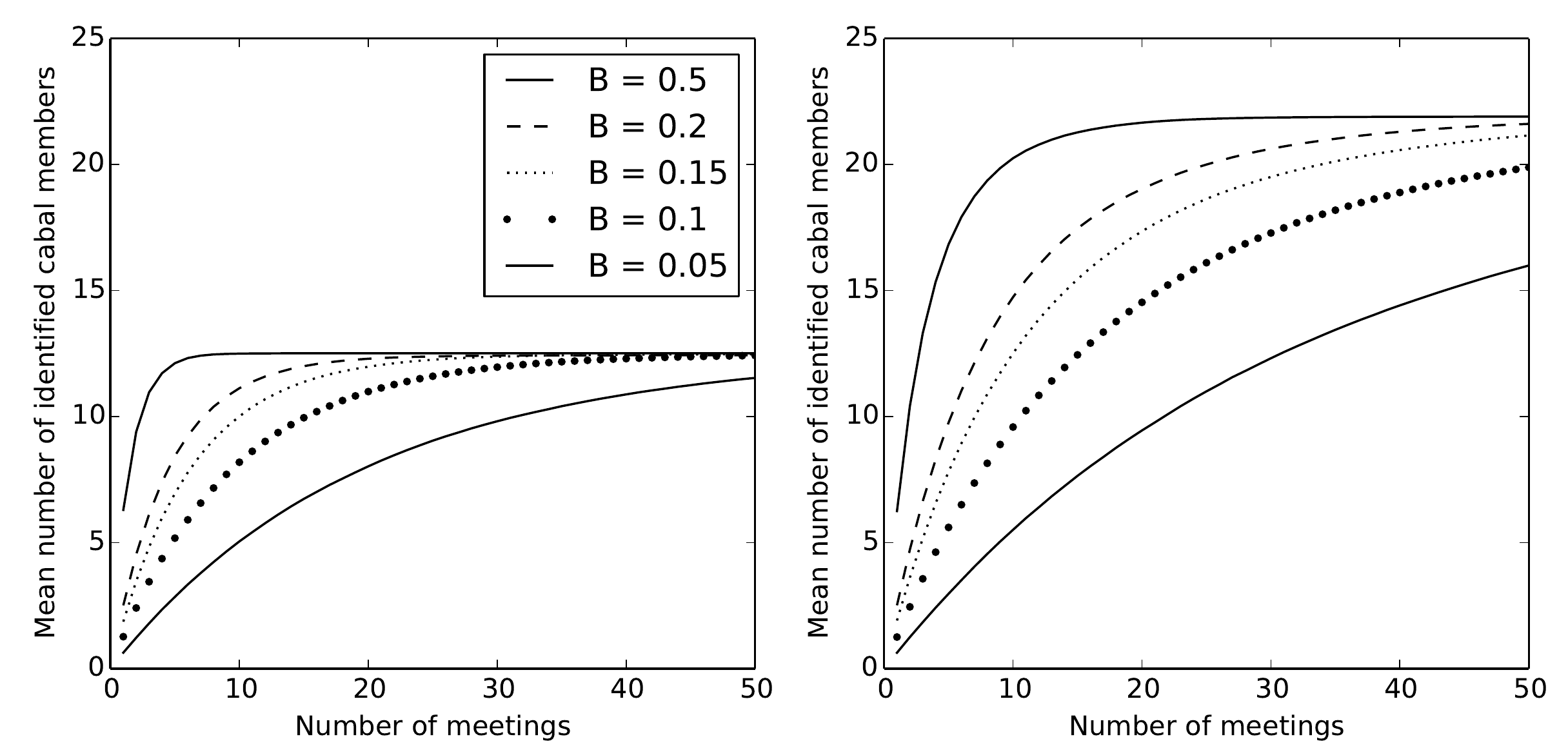}
\caption{Mean (over 10,000 trials) number of cabal members (out of 25) identified as a function of the number of cabal meetings for the one (left) and three (right) guards per client when the adversary controls $B=0.05$, $0.1$, $0.15$, $0.2$, and $0.5$ of the middle-relay bandwidth (different curves) and the probability of bridging a guard is $p_b = 0.5$.}\label{fig:mtor-sim-pb05}
\end{figure}

\begin{figure}
\includegraphics[width=0.45\textwidth]{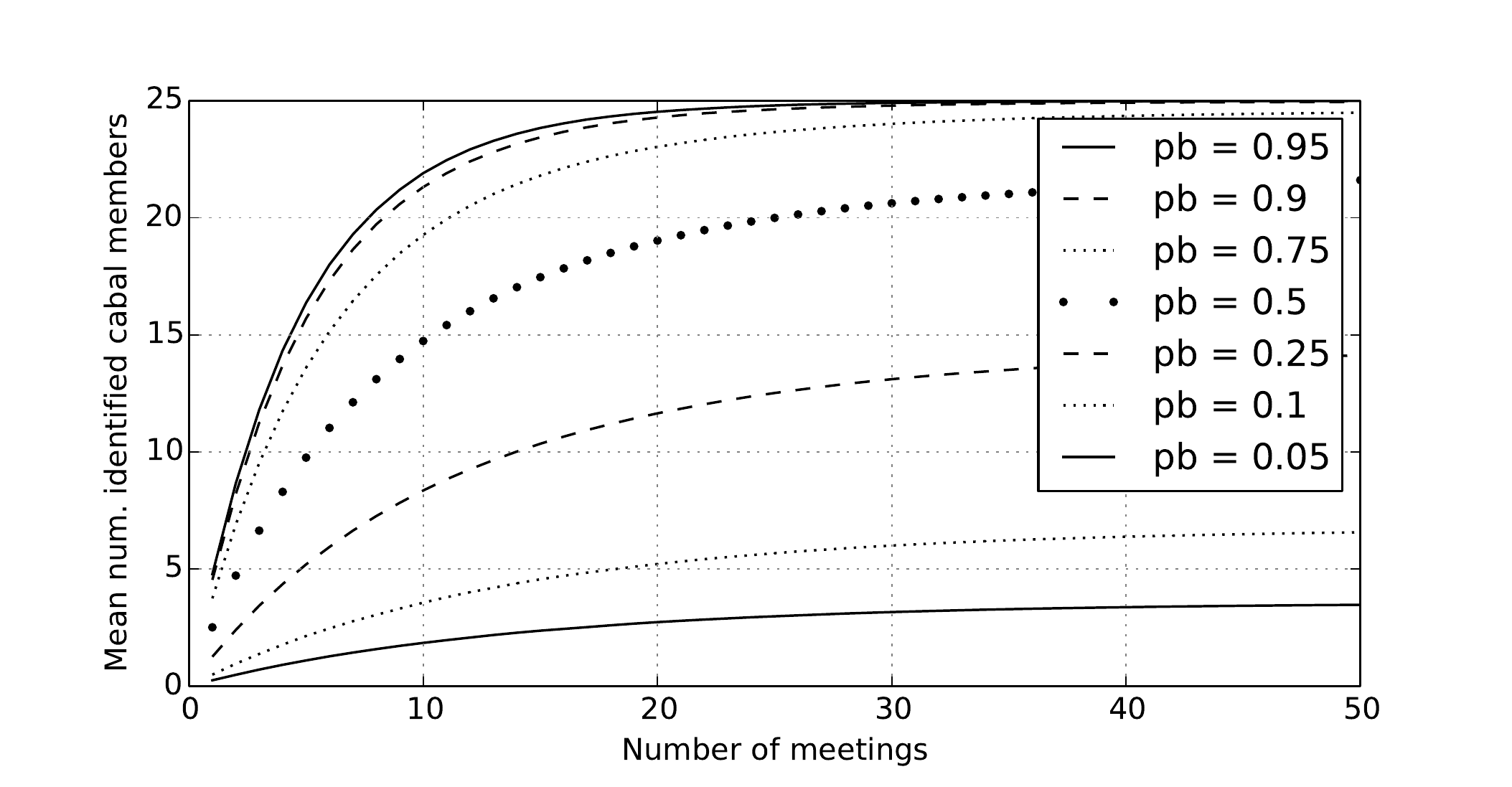}
\caption{Mean (over 10,000 trials) number of cabal members (out of 25) identified as a function of the number of cabal meetings in the three-guard case when the adversary controls $B=0.2$ of the middle-relay bandwidth and the probability of bridging a guard is $p_b = 0.05$, $0.1$, $0.25$, $0.5$, $0.75$, $0.9$, and $0.95$.
}\label{fig:mtor-sim-guard-pb}
\end{figure}

\paragraph{Results}
Figure~\ref{fig:mtor-sim-pb05} shows the mean number of identified cabal members, out of $25$ and averaged over 10,000 random trials, for different values of $B$ as a function of the number of meetings when each client has one (left) and three (right) guards.  This assumes a bridging probability $p_b = 0.5$.  Figure~\ref{fig:mtor-sim-guard-pb} is similar and shows the effects of the value of the bridging probability on these computations in the three-guard-per-client case.  The different curves in this plot correspond to different values of $p_b$; this assumes $B=0.2$.

Both $B$ and $p_b$ can have a significant impact on the adversary's success.  Figure~\ref{fig:mtor-sim-pb05} illustrates the benefit to the adversary of having additional guards that he may attempt to bridge (for a fixed $p_b$).  As we also discuss  below, the adversary might be able to increase $p_b$ against long-lived guards by continuing to devote resources to bridging them if initial attempts fail.  Figure~\ref{fig:mtor-sim-guard-pb} highlights the benefits of increasing $p_b$ by this or other means.

%% file: irc.tex
\section{Example: Targeting IRC Cabals}\label{sec:irc}

We now consider the following scenario: A cabal of interest to the adversary communicates via a private IRC channel.  All of the cabal members access the IRC server via Tor, and each creates a new Tor circuit for each cabal meeting.  In doing so, we assume a client chooses a guard uniformly at random from her set of guards and then chooses a middle relay uniformly at random from all middle relays.  The adversary compromises the middle relays independently with probability $B$, corresponding to the fraction of middle-relay bandwidth that he controls.  We assume that the network is static.

\subsection{Identifying guards and cabal members}

With respect to learning guards of cabal members, the analysis of
Sec.~\ref{ssec:mtor-guard} applies in this setting as well. The
assumption made there that probabilities for clients never
using a compromised middle relay are independent is even more
realistic here.

We also consider the adversary's success in identifying and locating
particular cabal members.  If the adversary already has a cabal
membership, by the properties of IRC he has a list of channel
pseudonyms for all cabal members, even those attending meetings
silently, and has a complete pseudonmymous record of all
communications. We again assume the adversary owns some fraction $B$
of the middle-relay bandwidth.  If the cabal leader Alice uses a
middle relay controlled by the adversary Sam, he observes the traffic
pattern through that relay and the matching messages that appears in
the channel and thus learns the guard used by Alice for that circuit.
Once the adversary knows this guard, he is able to bridge the guard
and identify Alice's IP address with probability
$p_b$. Other than through this combination of events, we assume the adversary
is not able to identify Alice's IP address.  Even if Sam only owns the
ISP of some targeted, uncompromised cabal member, he still passively
observes everything there, including the traffic pattern for a cabal's
IRC channel; as noted above, we assume that this or other information
allows the adversary to identify cabal traffic at middle relays.

We make the simplifying assumption that bridging is
actually with respect to client--guard pairs rather than individual
guards.  Thus, if clients $c_1$ and $c_2$ use the same guard $g$ for
circuits that go through compromised middle nodes (which may or may
not be the same for the two clients), then the adversary bridges $g$
and learns $c_1$ with probability $p_b$ and, \emph{independently},
bridges $g$ and learns $c_2$ with probability $p_b$.  Bridging that
arises from compromising the guard itself would not be independent for
these two clients, while bridging that arises from compromising client
ISPs or some part of the network between the clients and $g$ might be.
We thus think this assumption is reasonable, although others could be made as
well.

\begin{figure}[htb]
\includegraphics[width=0.45\textwidth]{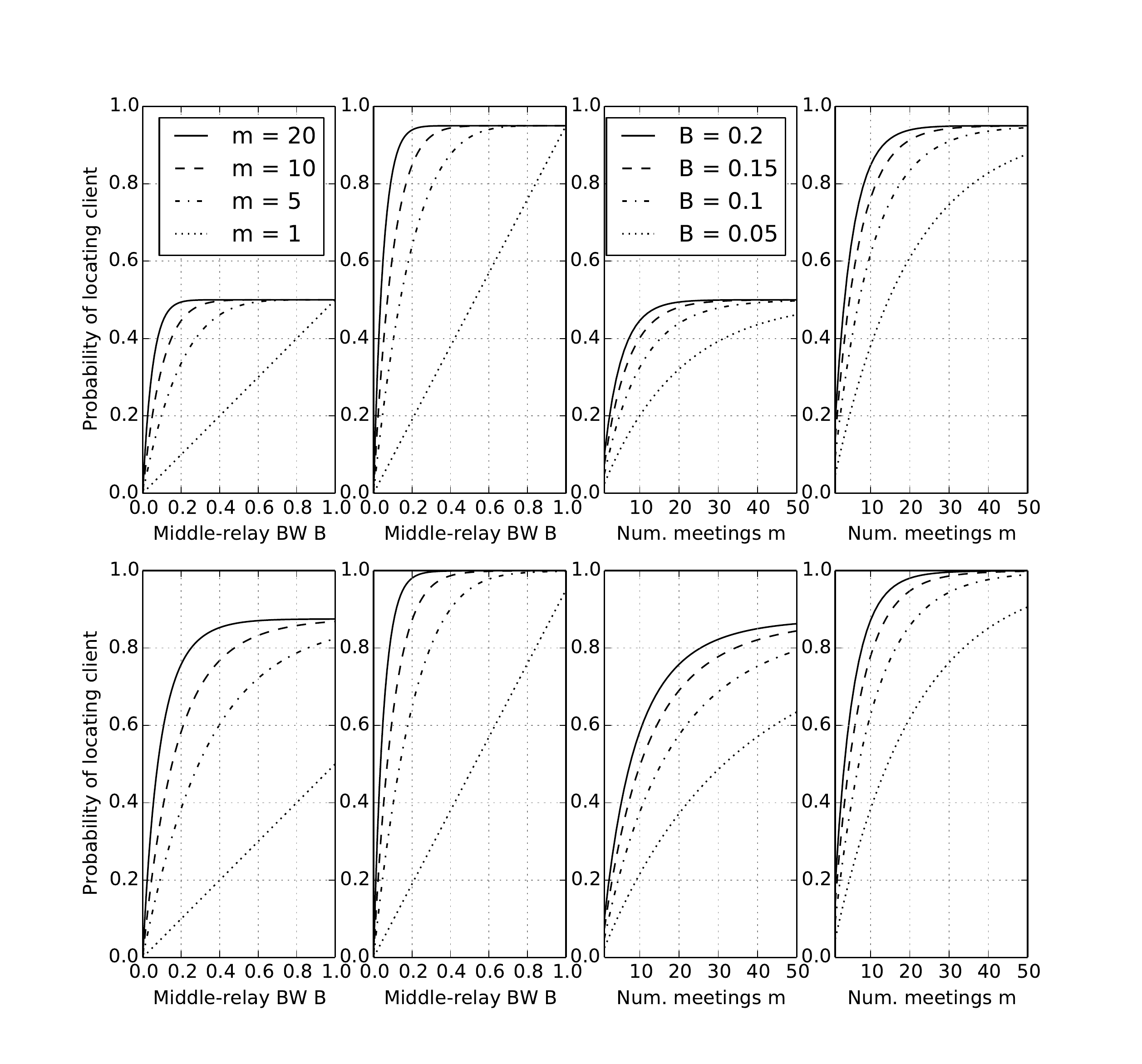}
\caption{Probability that the attacker is able to identify the cabal leader for one (top row) and three (bottom row) guards per cleint when the attacker controls a fraction $B$ of the middle-relay capacity over the course of $m$ meetings.  This is shown as a function of $B$ in the left column (with different curves for different values of $m$) and as a function $m$ in the right column (with different curves for different values of $B$).  The bridging probability $p_b$ is, in each quadrant, $0.5$ (left subplot) and $0.95$ (right subplot).}
\label{fig:irc-ident}
\end{figure}

Figure~\ref{fig:irc-ident} plots the probabilities, for various cases, that the adversary is able to identify the cabal leader.  The computations are discussed in App.~\ref{app:irccomp}.  The top row assumes one guard per client; the bottom row assumes three guards per client.  The left column (pair of subplots) shows plots this probability as a function of $B$ for $m$ meetings (different curves for $m = 1$, $5$, $10$, and $20$).  The right column (pair of subplots) shows this as a function of $m$ with different curves for $B = 0.05$, $0.1$, $0.15$, and $0.2$.  Each pair of subplots shows bridging probability $p_b$ of $0.5$ (left) and $0.95$ (right).

Considering Fig.~\ref{fig:irc-ident}, the attacker is likely to be able to identify the cabal leader if he makes a substantial but not unrealistic investment in middle-relay bandwidth ($B=0.2$), even if the cabal has a modest number of meetings (10 or more).  The left subplot in the top-left pair shows the adversary's chances bounded by $p_b=0.5$ because there is only one guard to bridge.  However, we expect that, if there is a single long-lived guard, the adversary's chance of bridging the guard would go up over time, and the success probabilities would become closer to those shown in the right subplot in this pair.

Separately, we note that (ignoring the fact that the attacker cannot expect to kill every circuit) killing one circuit per client per meeting would compress the $m$ axis in Fig.~\ref{fig:mtor-comp-fraction} (which, as noted, applies to the IRC case as well) by a factor of $2$.  If the attacker were willing and able to kill $k$ total circuits per client, it could shift these curves to the left by $k$.  Considering Fig.~\ref{fig:mtor-comp-fraction}, we see that the greatest benefit to the attacker comes in the first circuit that he kills.

\subsection{Estimating cabal size}

We turn now to estimating the total size of the cabal (when the
adversary does not already own a member).  In particular, if an IRC cabal
of size $c$ has $m$ meetings, what estimate of $c$ will be made by an
adversary who controls a fraction $B$ of middle-relay bandwidth?  How
is this estimate likely to be distributed, and how much of an error is
he likely to make?

To answer these questions, we numerically compute the maximum-likelihood estimator (MLE) of the cabal size based on an $m$-tuple $\vec{x} \in \{0,1,\ldots,c\}^m$ of observations that the adversary might make of the number of cabal members using compromised middle relays during each of $m$ meetings.  We then compute the distribution on possible observations to determine, for each estimated value, the probability that the adversary will make observations that give rise to that estimate.

In doing this, we assume that the adversary considers only the number of circuits that he sees during each meeting window and not the guards used for these circuits.  In the case that each client uses only one guard, if the adversary sees one circuit in each of three meetings and these circuits all use different guards, then he knows that he has \emph{observed} three different clients and not just one client multiple times.  As a result, he should probably increase his estimate of the cabal size compared to his estimate when simply using a count of the circuits he observes.  The approach we take already requires significant computational resources, and accounting for guard identities makes it even more complex to the point that we expect it would be infeasible (while also not adding significantly to the adversary's accuracy).  We do note that this issue may have an effect.  (We compute the probabilities of guard collisions and show these in Tab.~\ref{tab:gcprob} in App.~\ref{ap:guardcoll}.)

\begin{figure}[htb]
\includegraphics[width=0.5\textwidth]{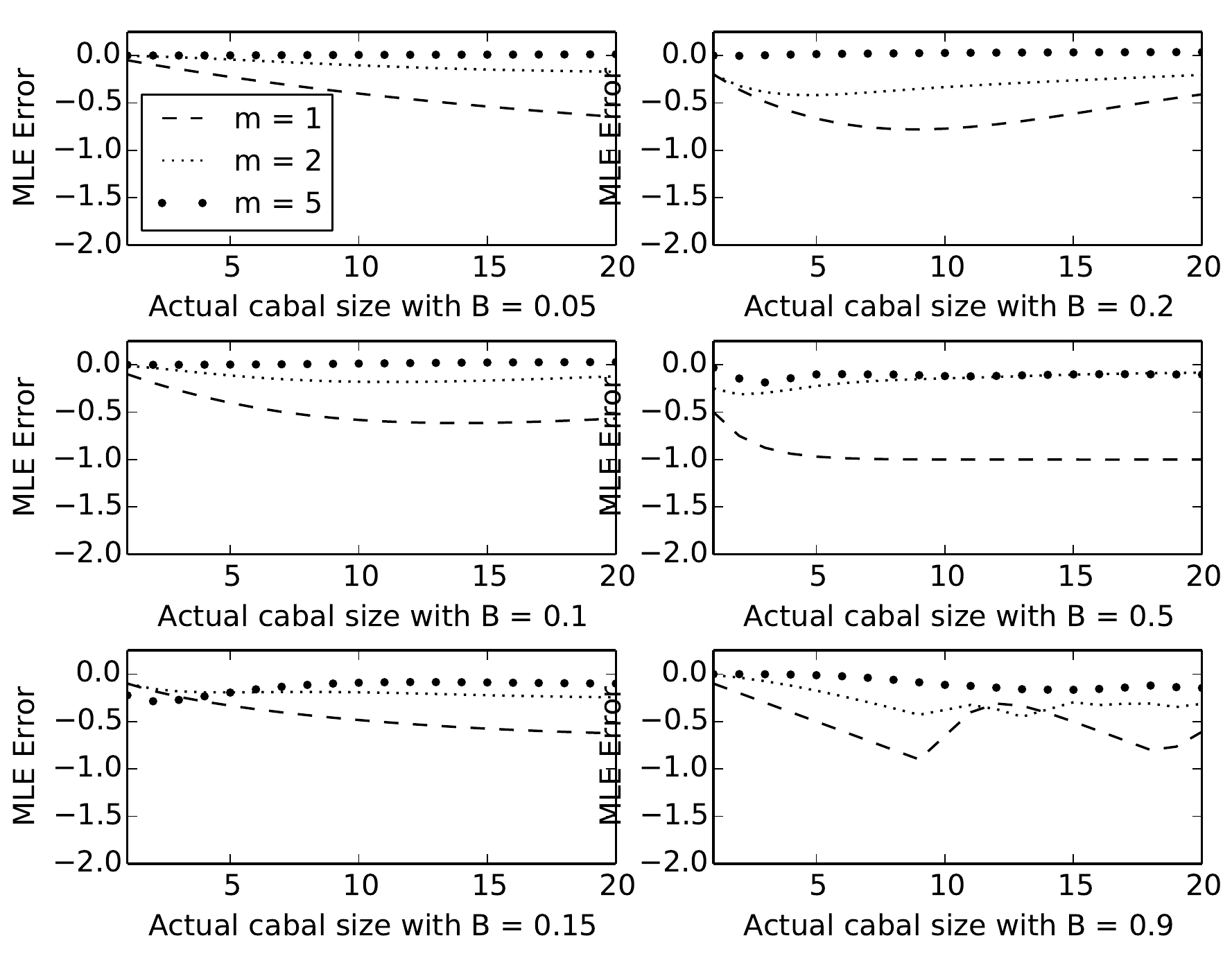}
\caption{Expected value of the MLE for the number $c$ of cabal members minus $c$ as a function of $c$ for number of meetings $m = 1$, $2$, and $5$ (different curves).  The left column has the adversary-controlled fraction of middle-relay bandwidth $B = 0.05$, $0.1$, and $0.15$ (top to bottom); the right column has $B = 0.2$, $0.5$, and $0.9$ (top to bottom).}
\label{fig:basic-emle}
\end{figure}

Figure~\ref{fig:basic-emle} shows the error in the expected value of the MLE---\ie, the expected MLE minus the actual cabal size $c$---as a function of $c$ for different values of $B$ (different subplots) and $m$ (different curves).

\begin{figure}
\includegraphics[width=0.12\textwidth]{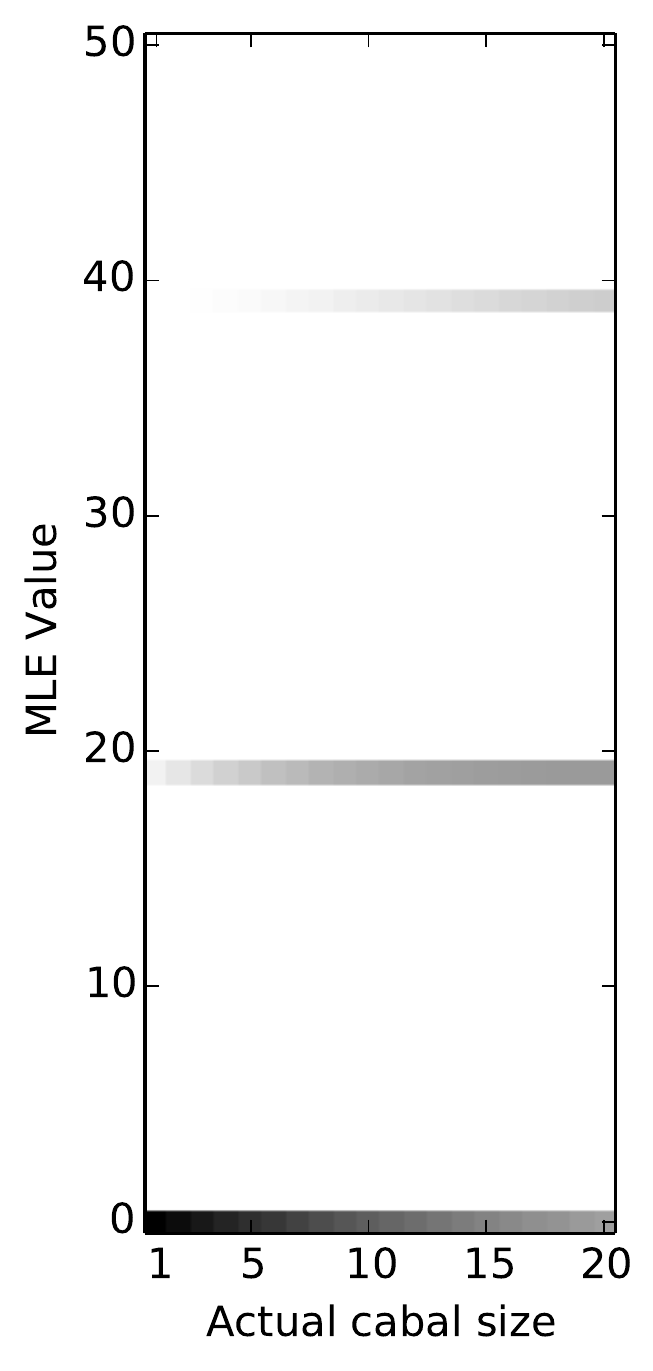}\includegraphics[width=0.12\textwidth]{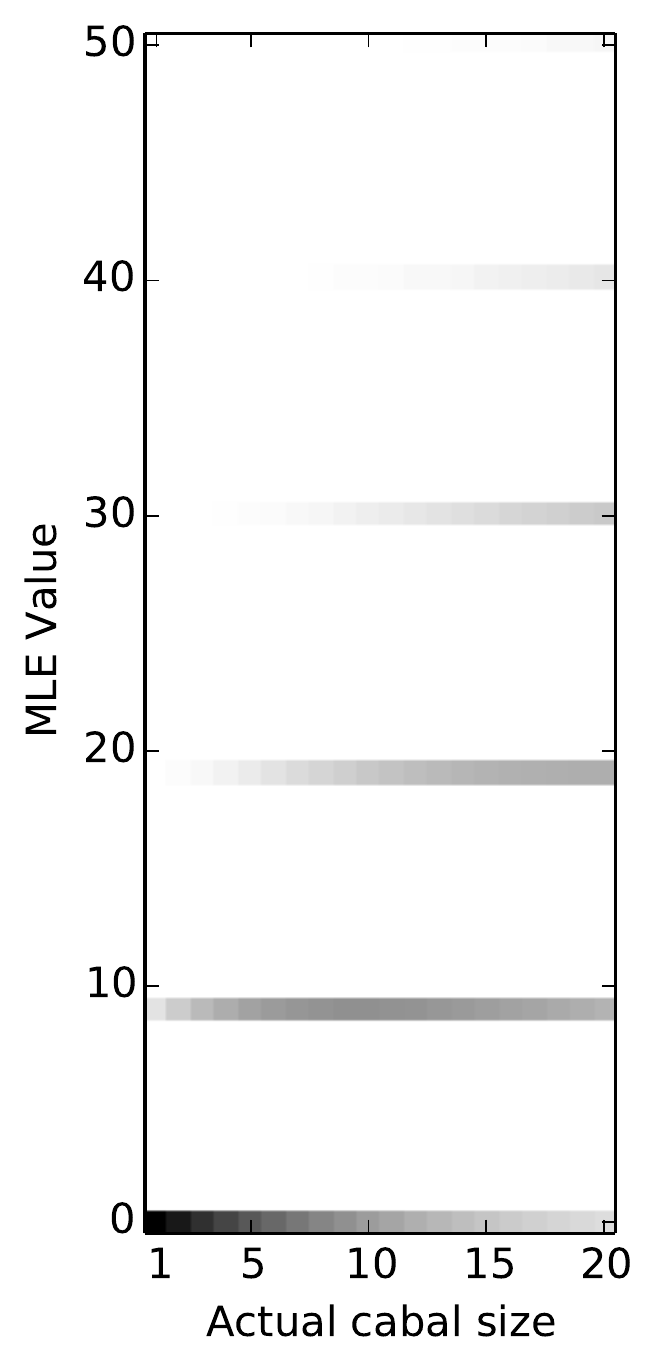}\includegraphics[width=0.12\textwidth]{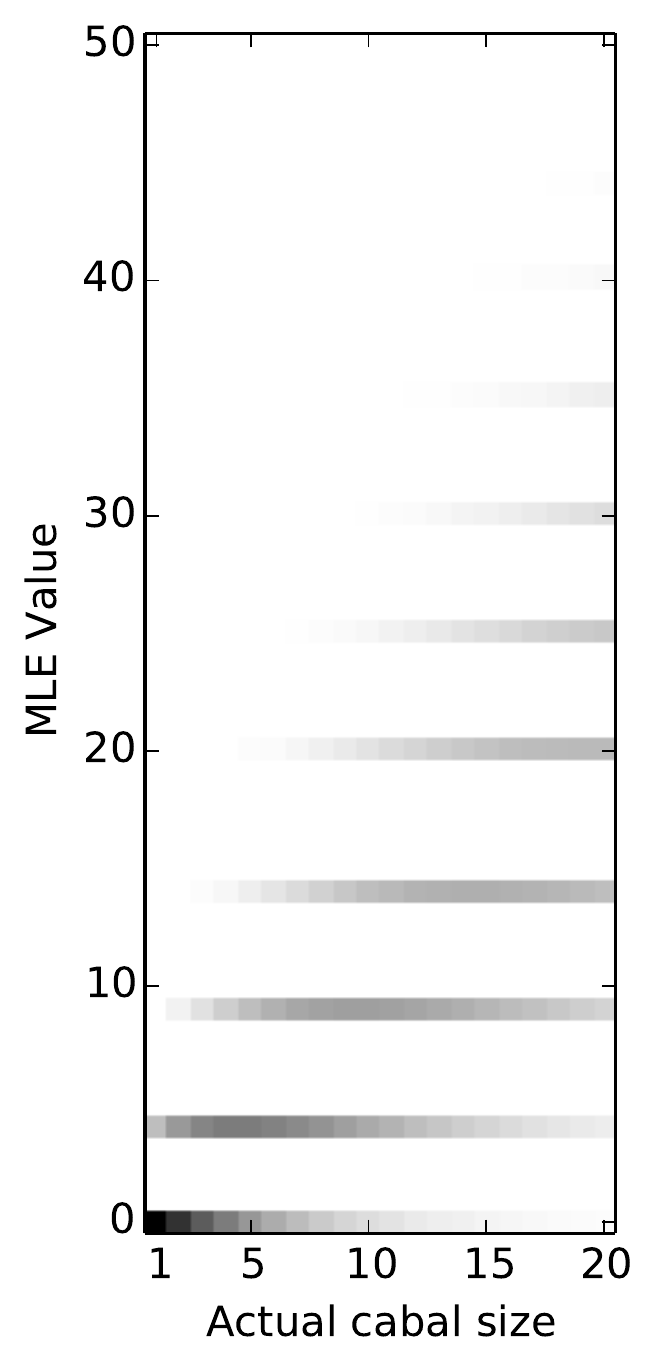}\includegraphics[width=0.12\textwidth]{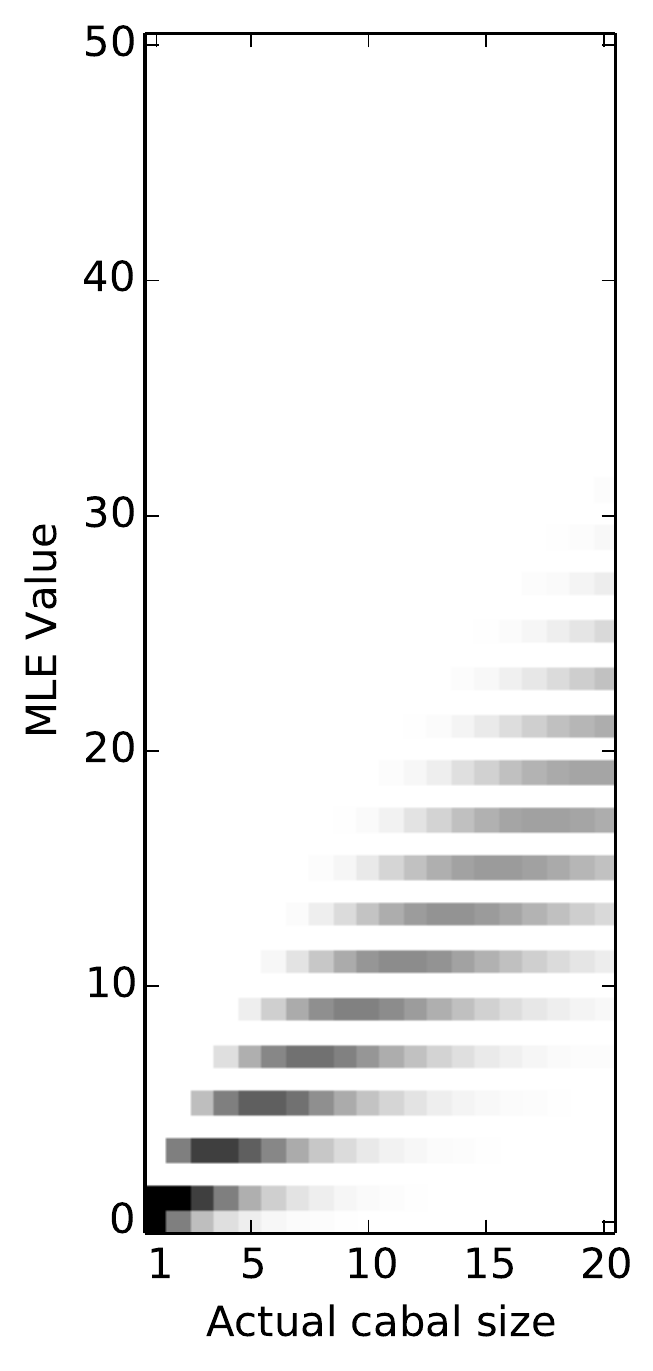}\\
\includegraphics[width=0.12\textwidth]{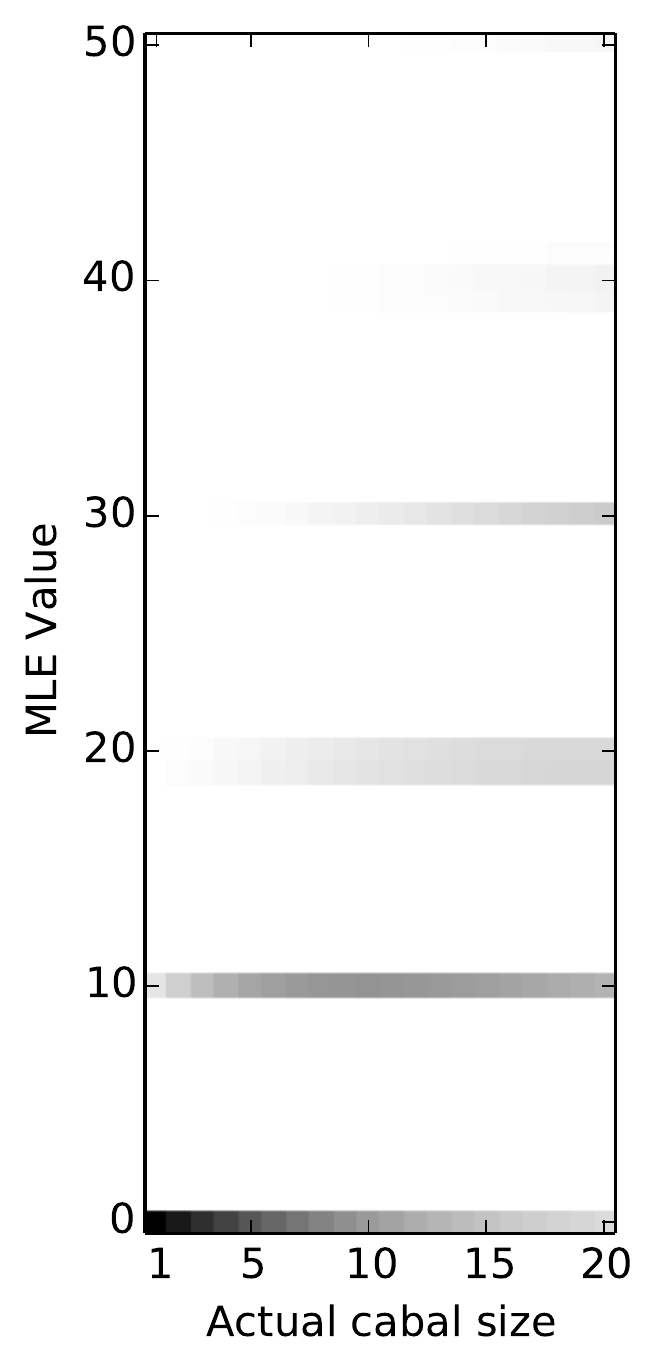}\includegraphics[width=0.12\textwidth]{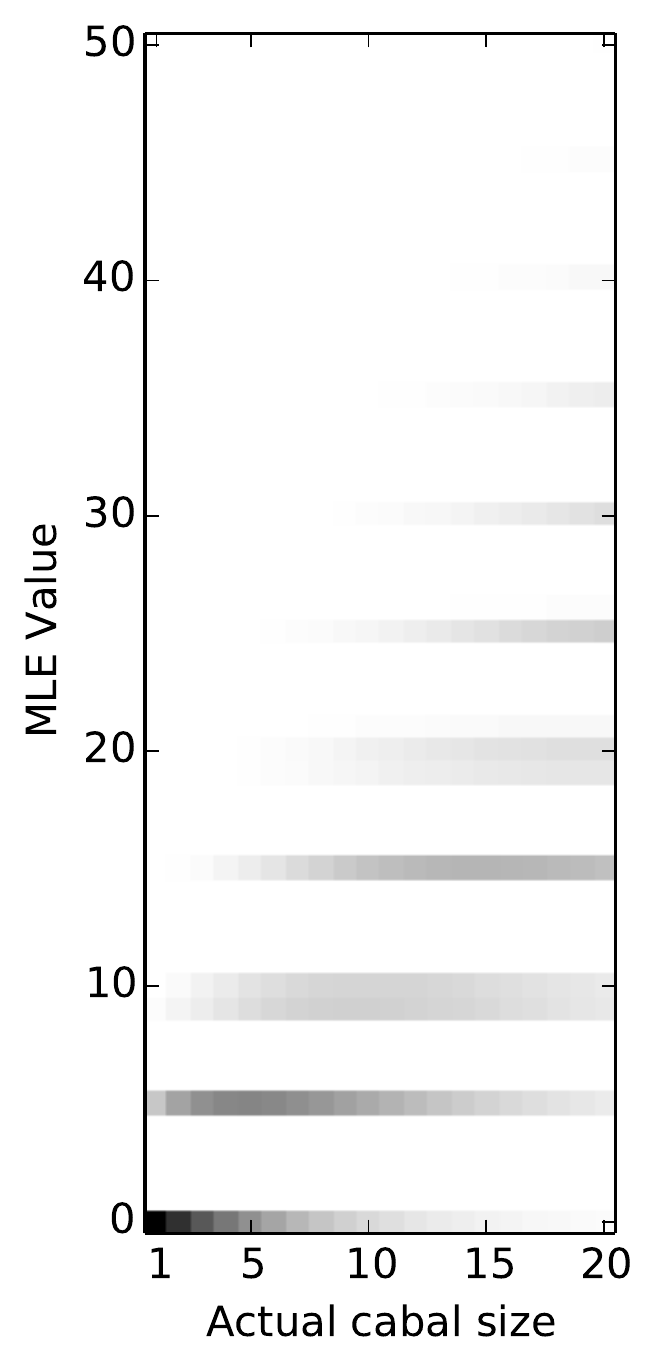}\includegraphics[width=0.12\textwidth]{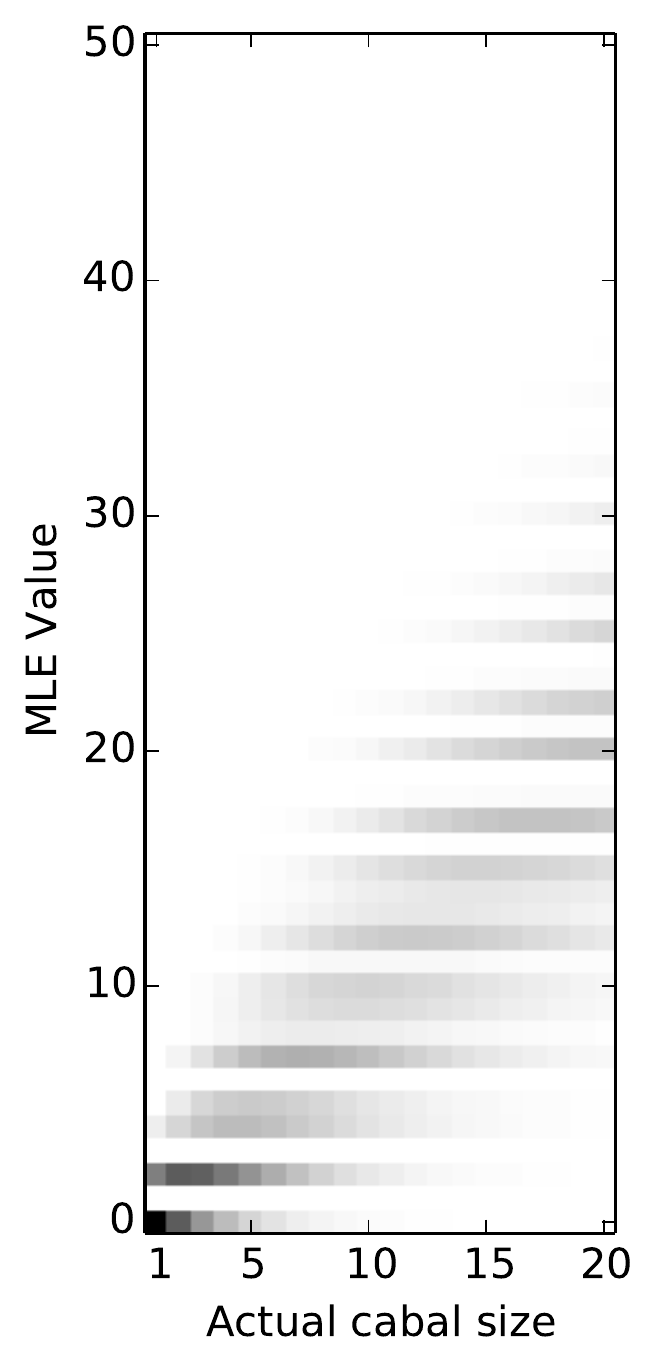}\includegraphics[width=0.12\textwidth]{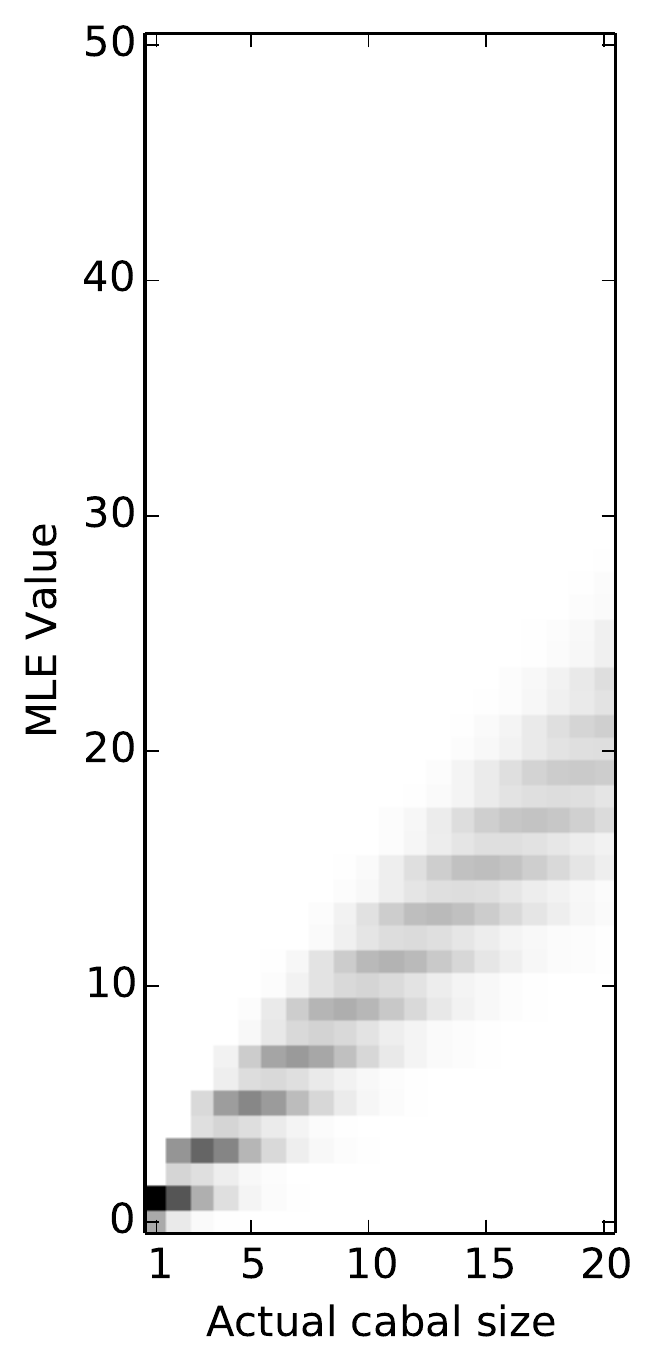}\\
\includegraphics[width=0.12\textwidth]{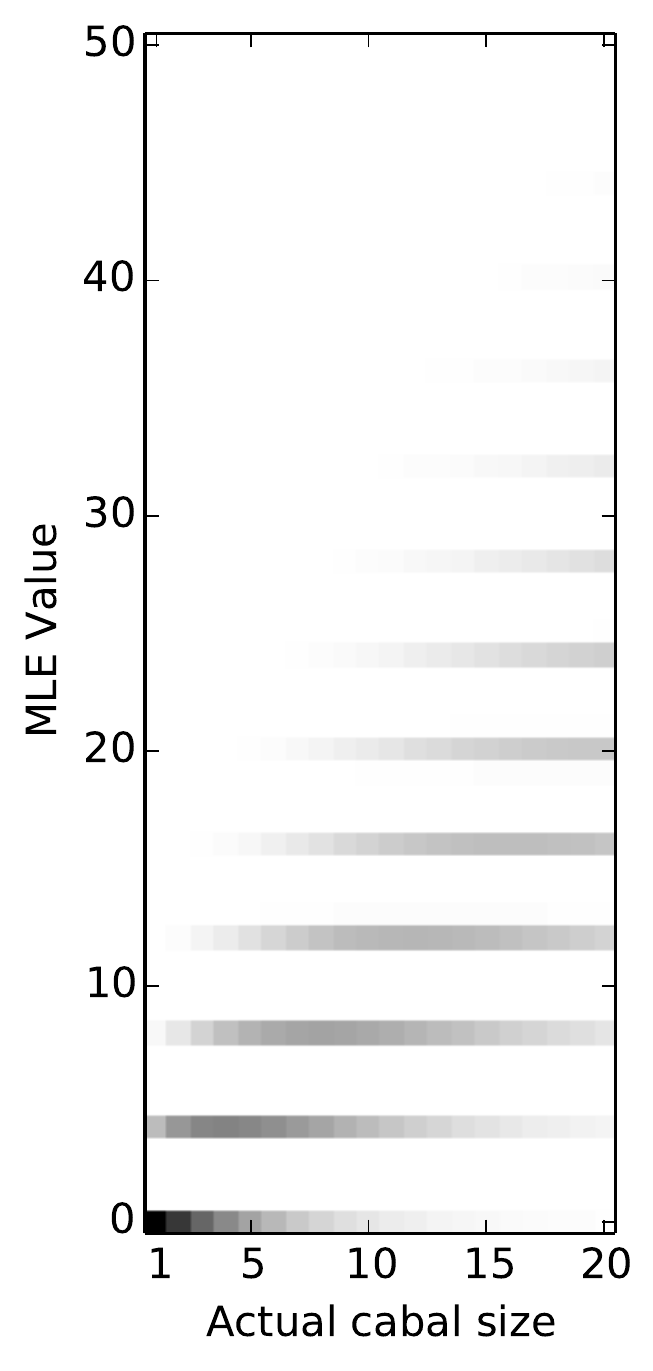}\includegraphics[width=0.12\textwidth]{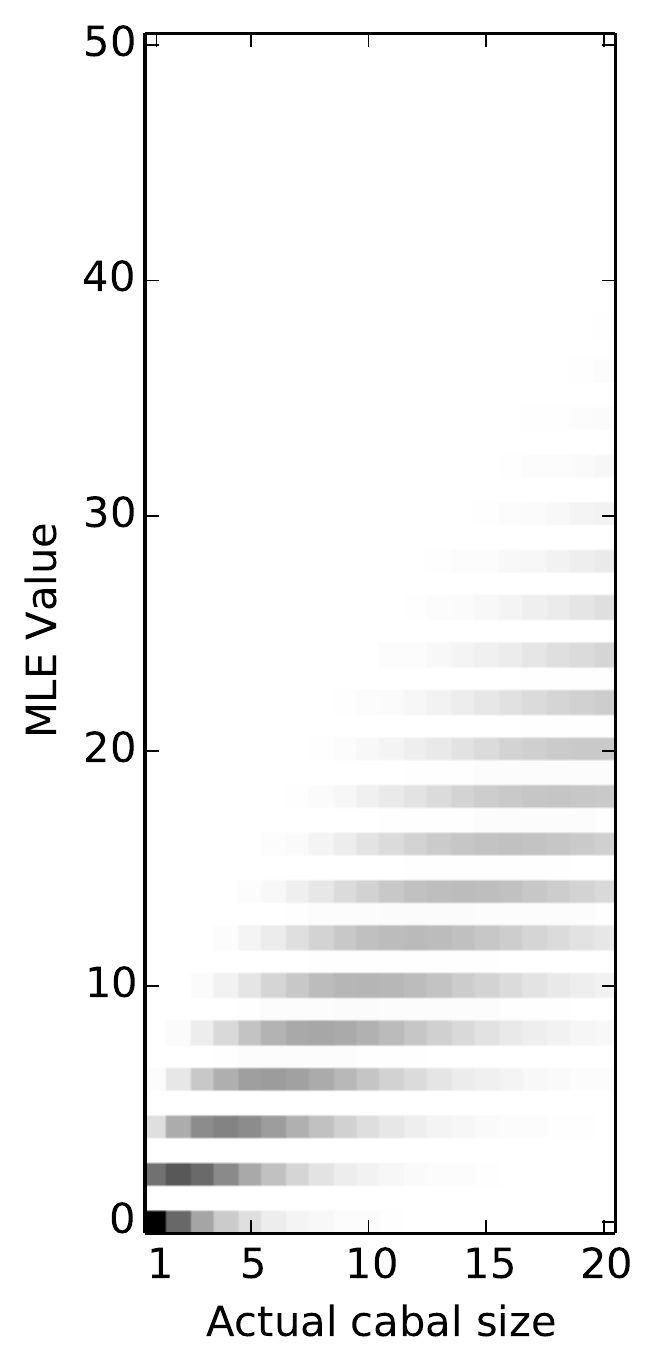}\includegraphics[width=0.12\textwidth]{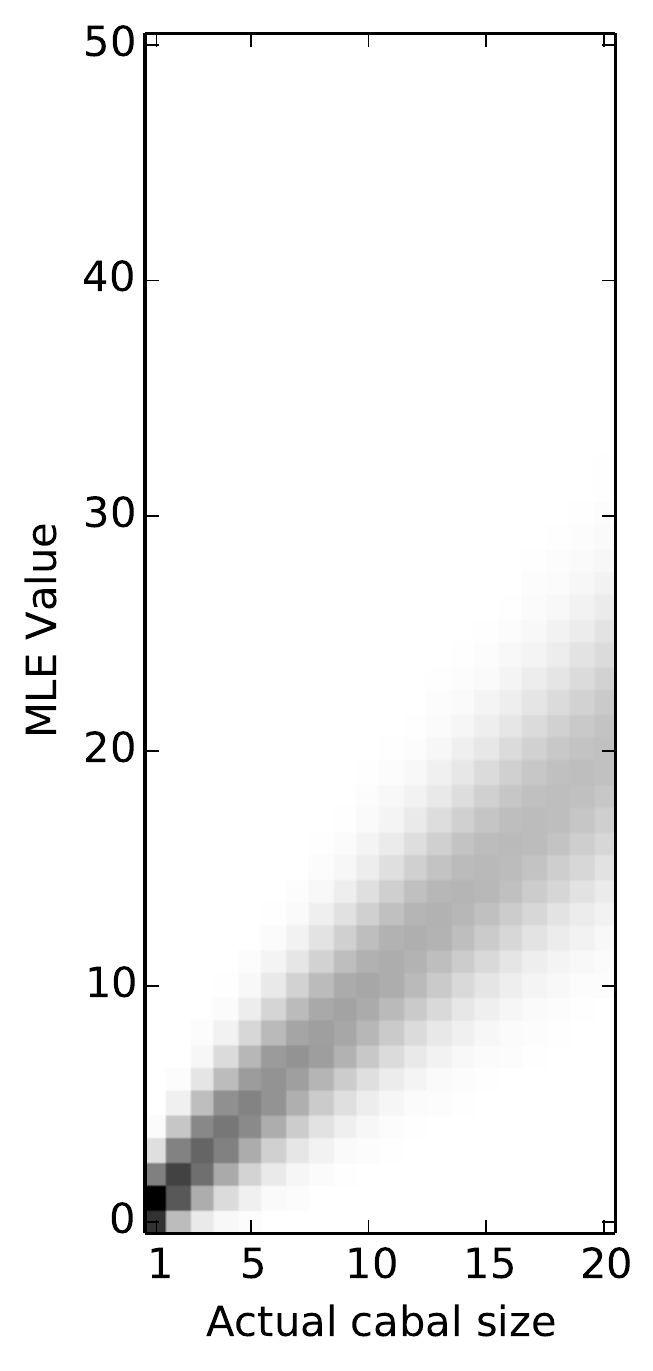}\includegraphics[width=0.12\textwidth]{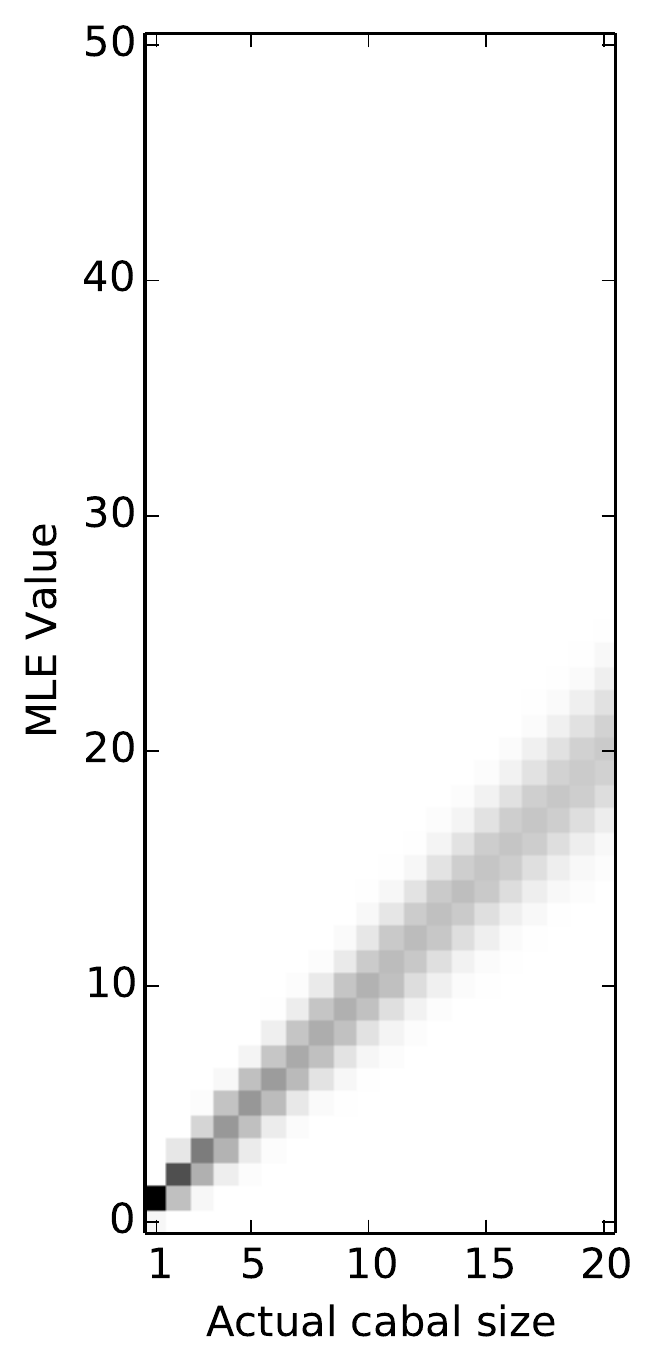}
\caption{The effect of increasing the number of meetings $m$ and changing the adversary-controlled middle-relay bandwidth $B$.  From top to bottom, $m = 1$, $2$, and $5$.  From left to right, $B = 0.05$, $0.1$, $0.2$, and $0.5$.  Within each subplot, the actual cabal size $c$ increases from $1$ to $20$ along the horizontal axis, and possible MLE values increase from $0$ to $50$ up the vertical axis.  For each $c$ and MLE value, the plot indicates the probability (darker values are larger) of the adversary observing a tuple for which it would compute the indicated value as its MLE for $c$.}\label{fig:dmle-single-matrix}
\end{figure}

Figure~\ref{fig:dmle-single-matrix} presents a matrix of plots of distributions of the MLE value.  Each subplot shows, for each value $c$ on the horizontal axis, the distribution of MLE values, with larger probabilities corresponding to darker shading.  In the matrix of plots, $B$ increases from left to right ($0.05$, $0.1$, $0.15$, $0.2$, $0.5$, $0.75$, and $0.9$), and $m$ increases from top to bottom ($1$, $2$, and $5$).

\begin{figure*}
\begin{centering}
\includegraphics[height=0.13\textheight]{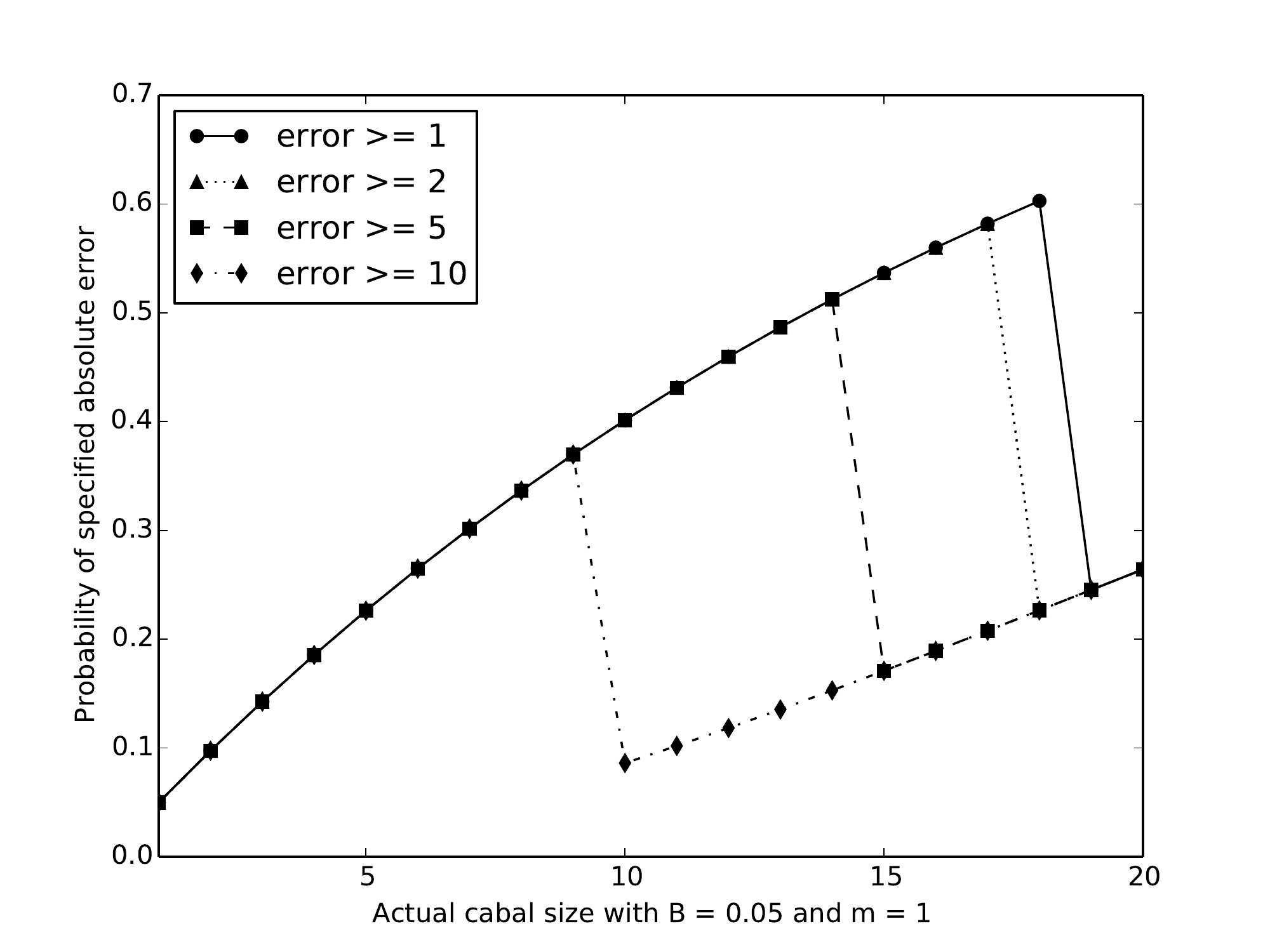}\includegraphics[height=0.13\textheight]{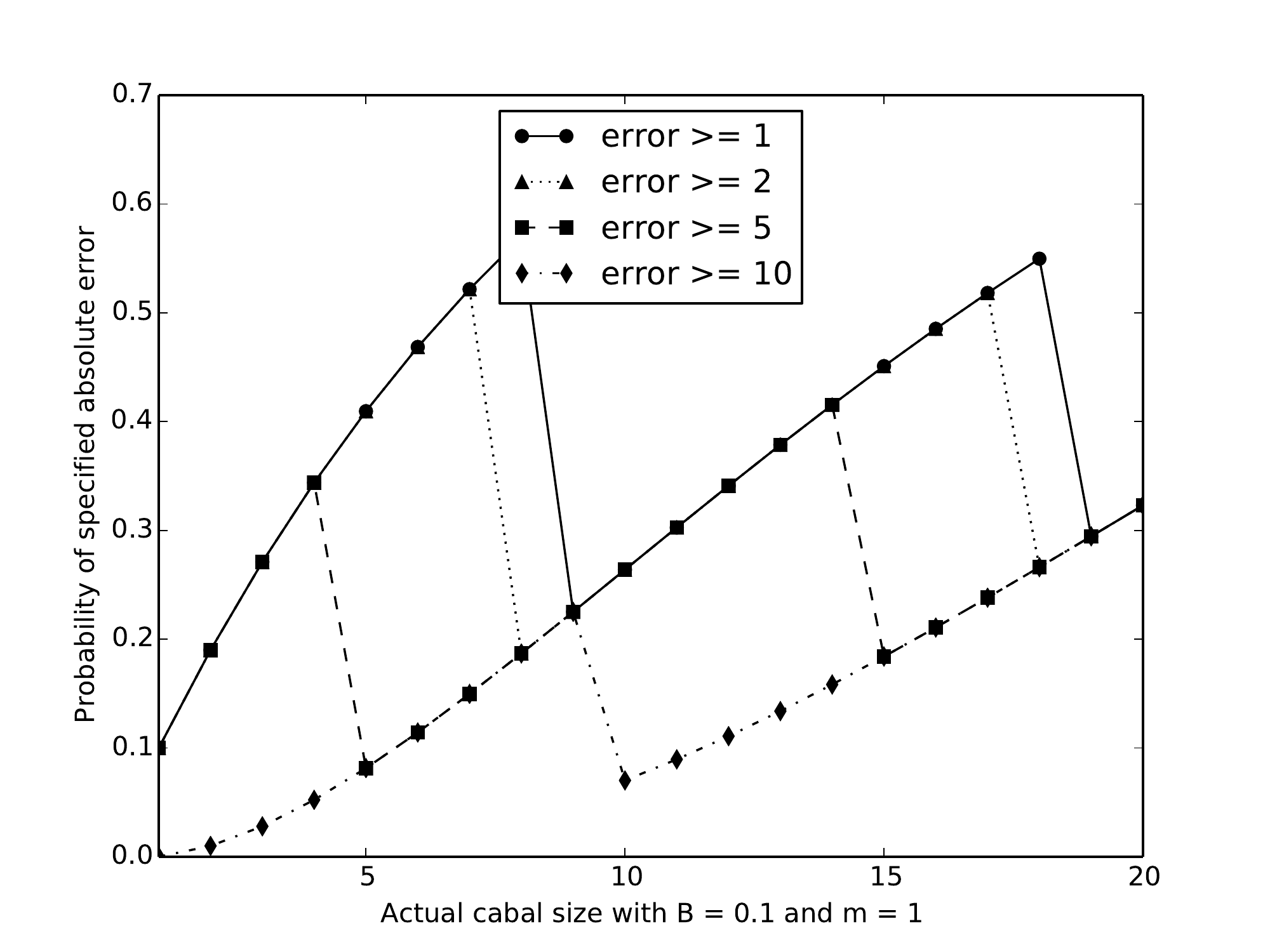}\includegraphics[height=0.13\textheight]{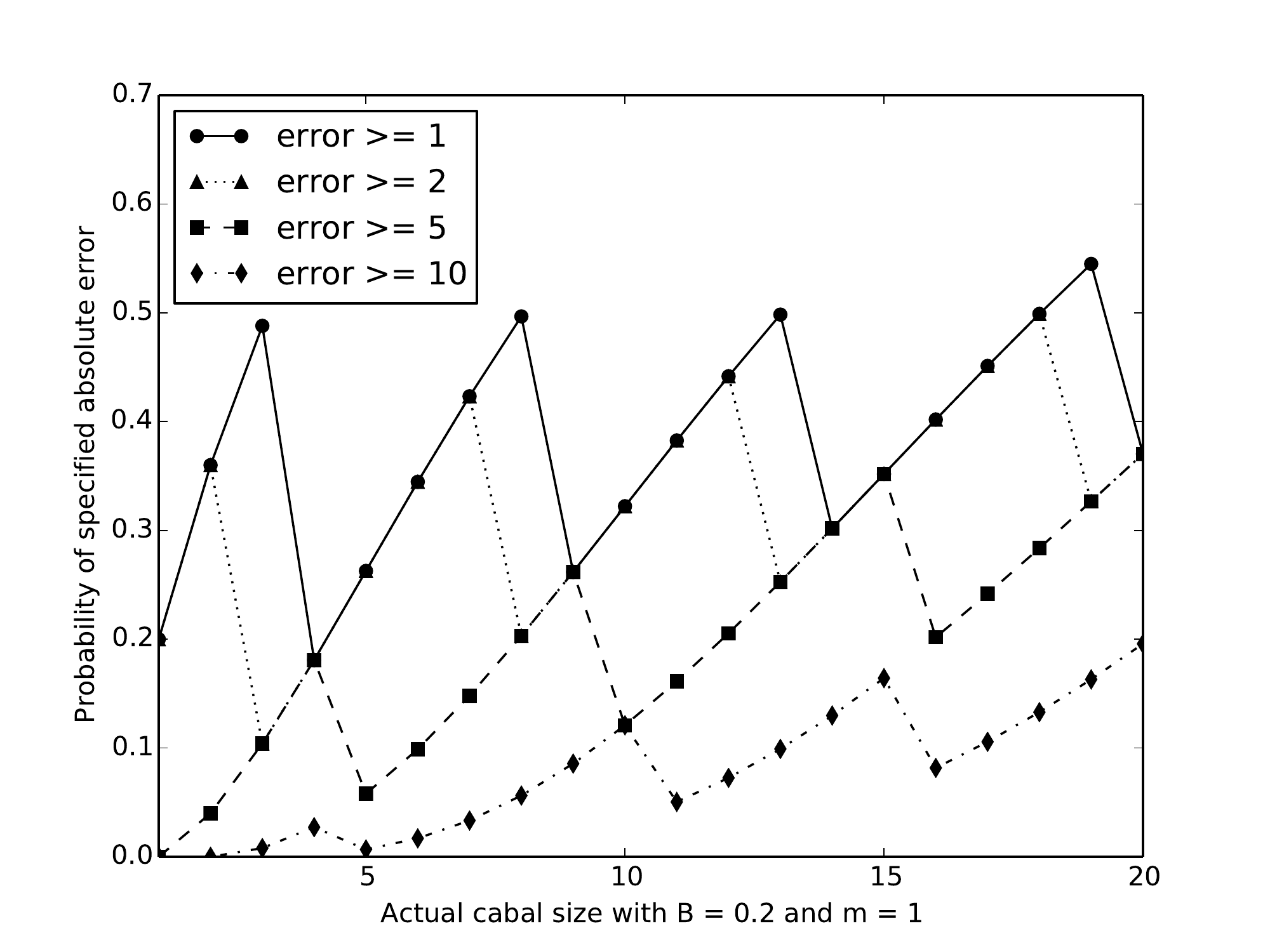}\includegraphics[height=0.13\textheight]{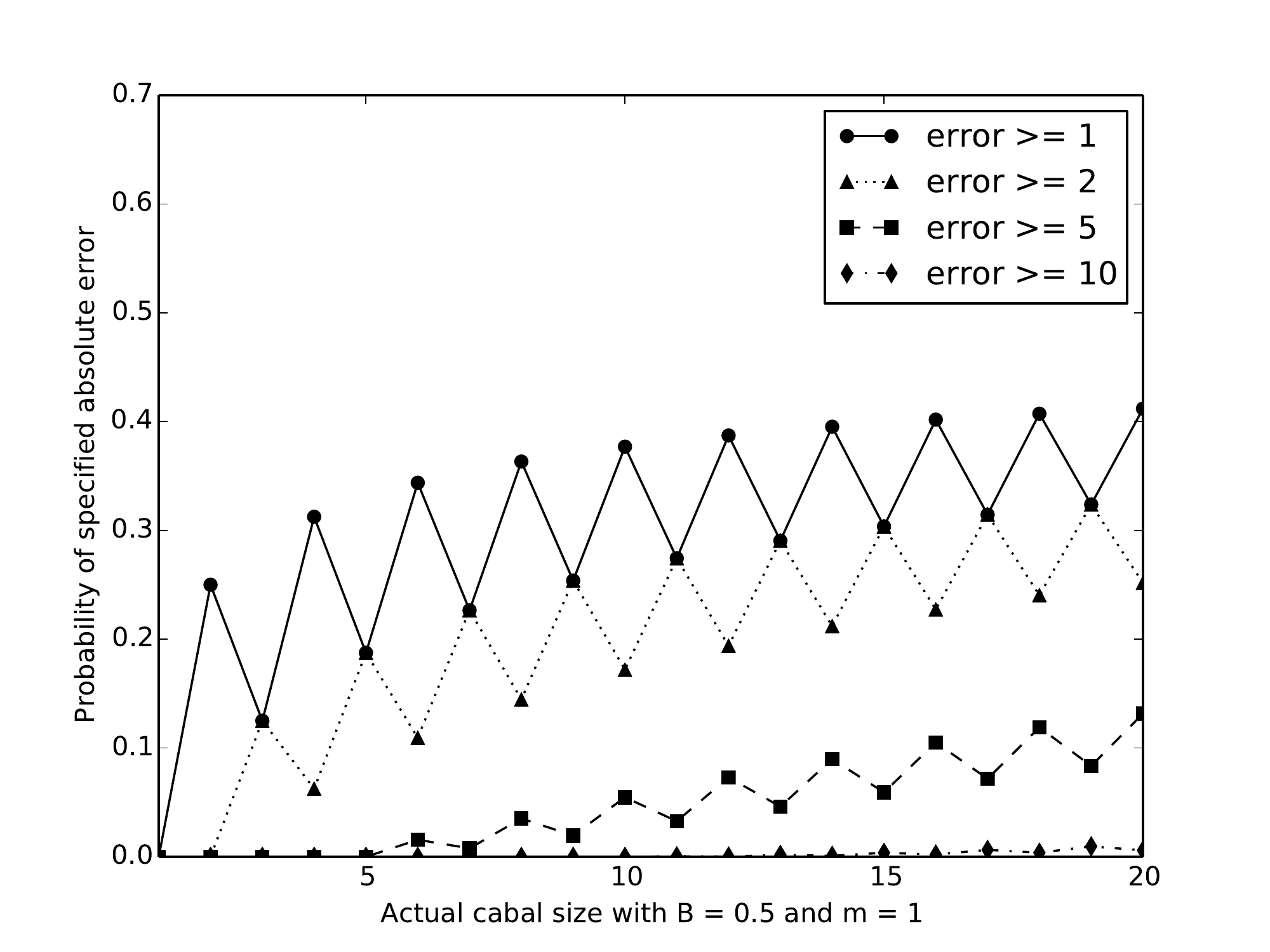}\\
\includegraphics[height=0.13\textheight]{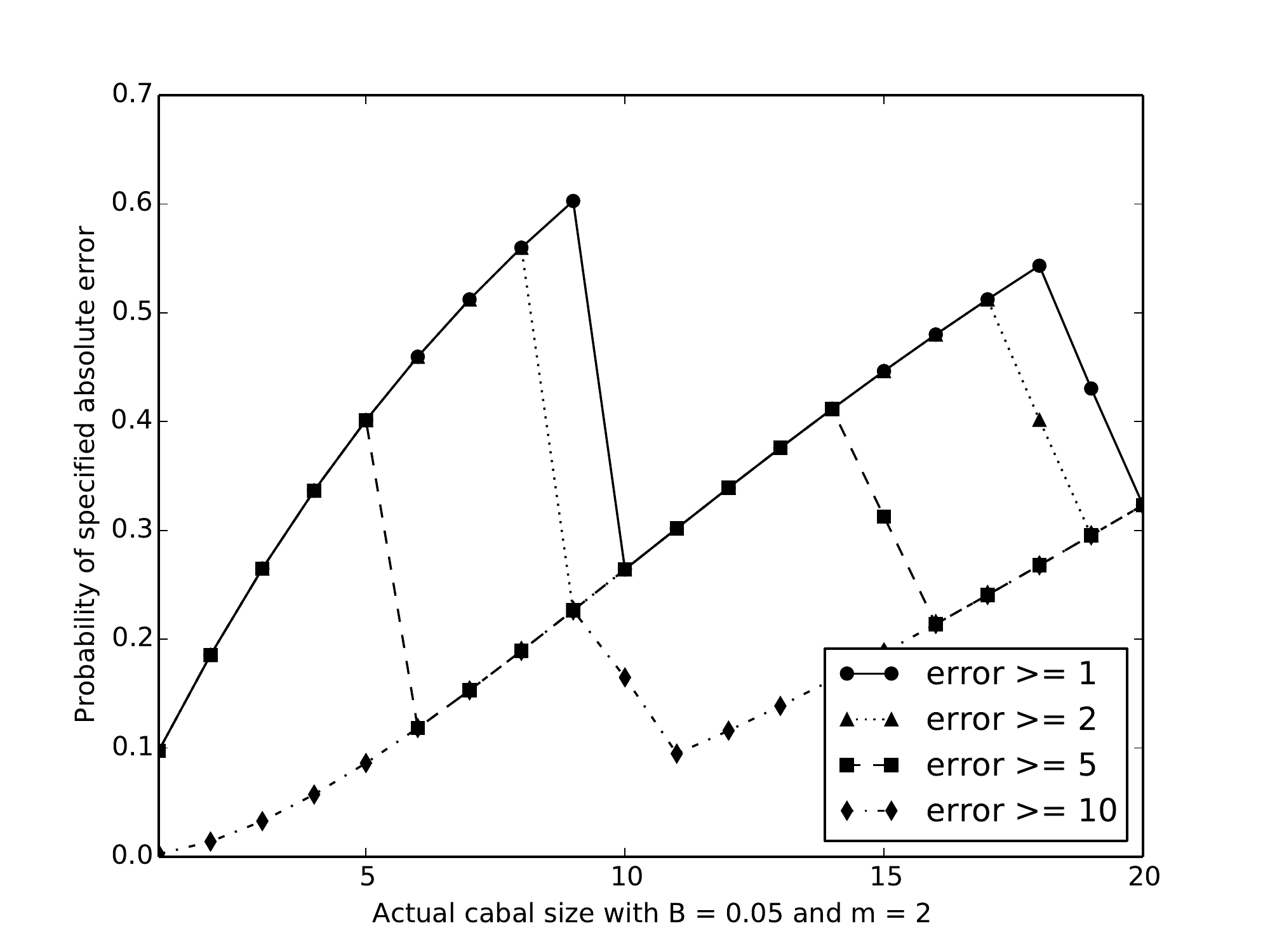}\includegraphics[height=0.13\textheight]{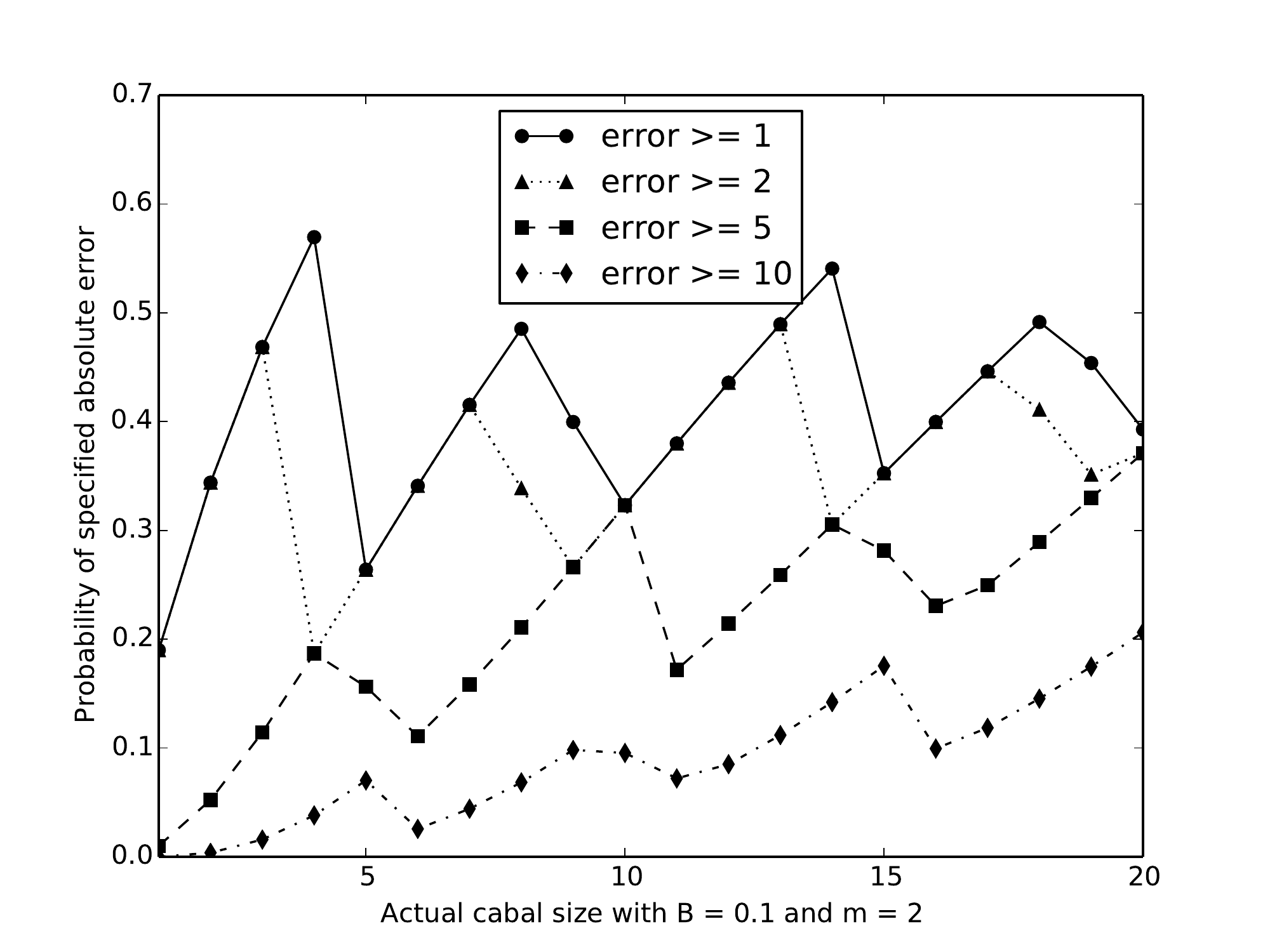}\includegraphics[height=0.13\textheight]{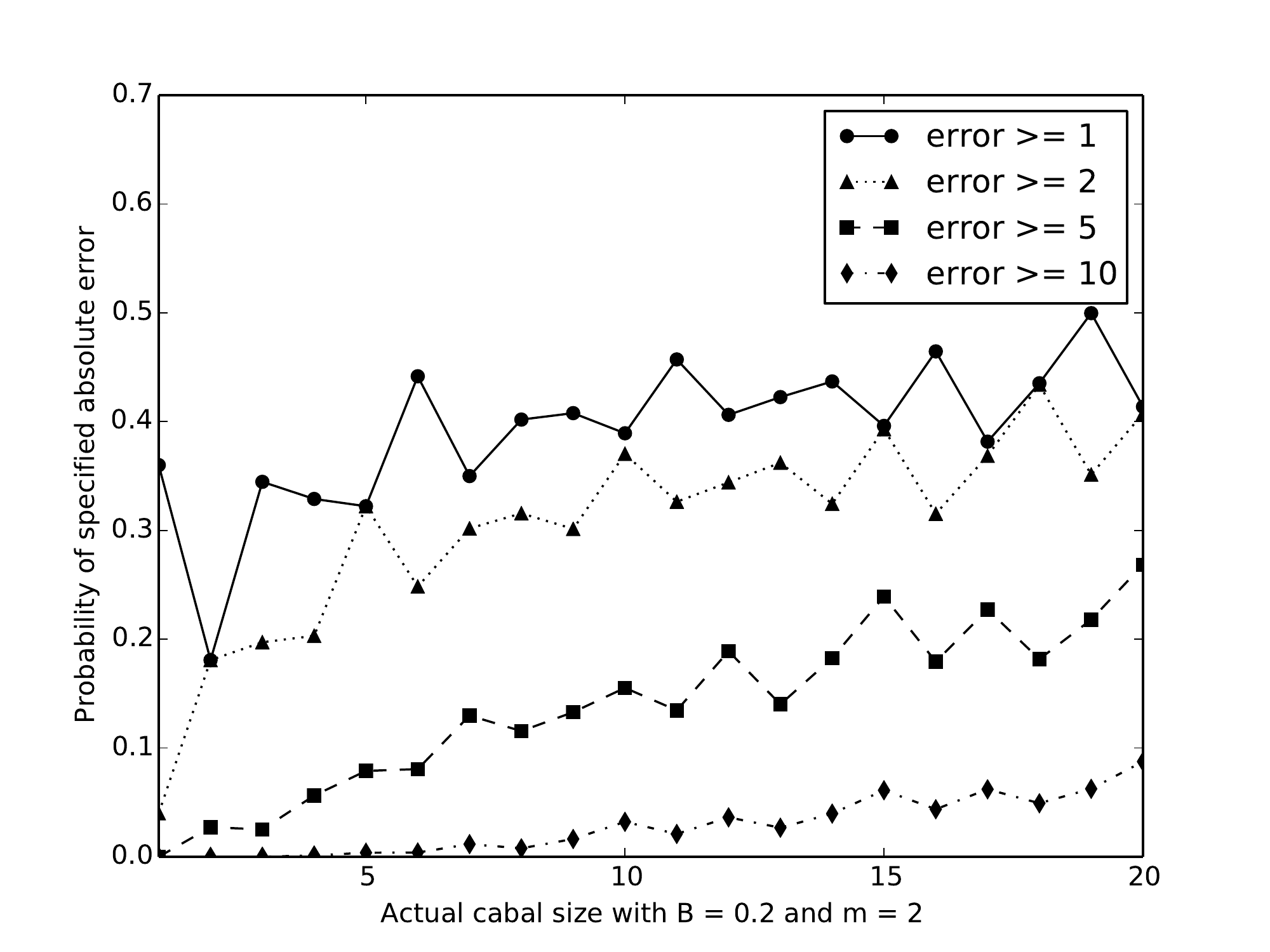}\includegraphics[height=0.13\textheight]{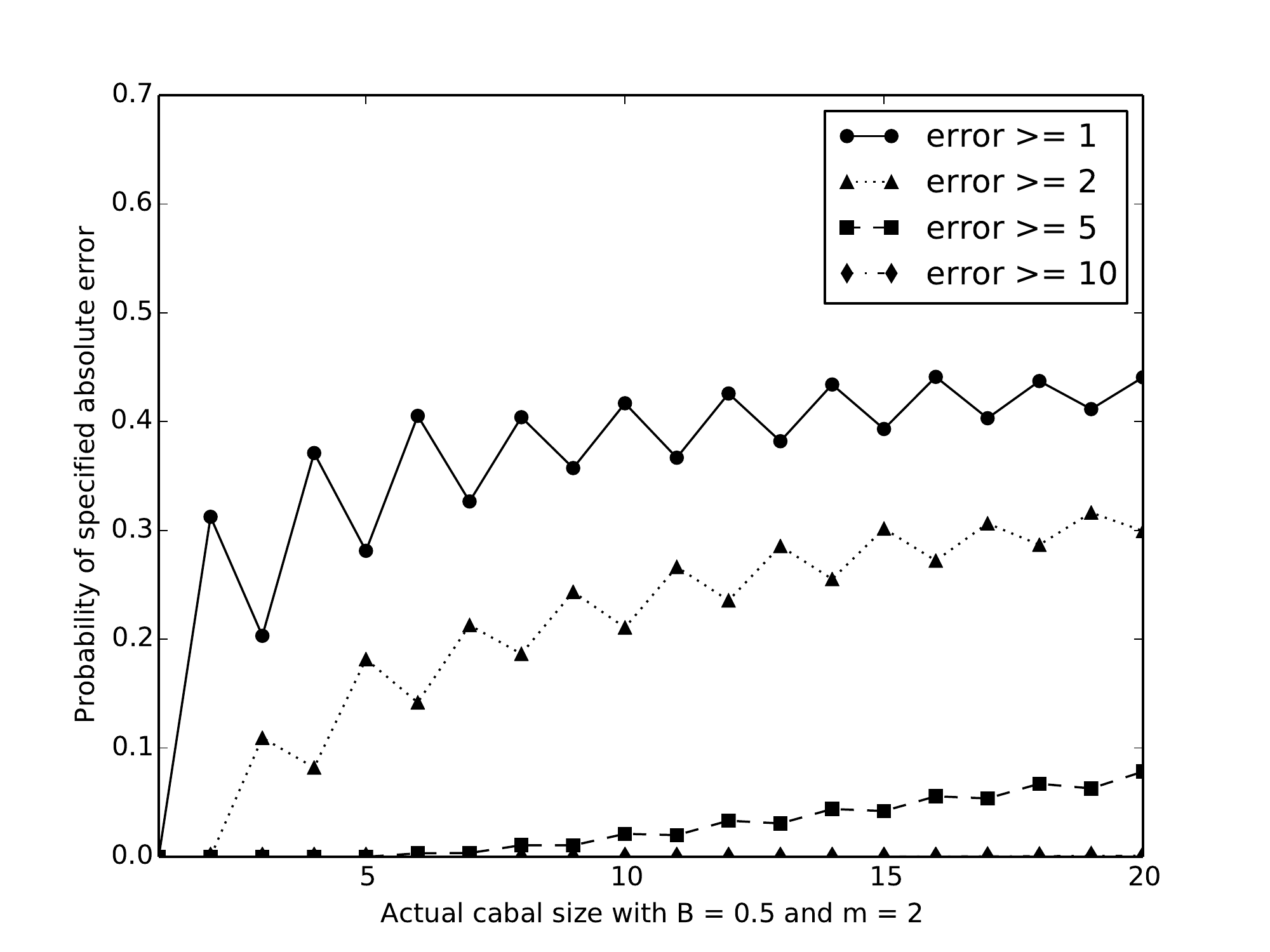}\\
\includegraphics[height=0.13\textheight]{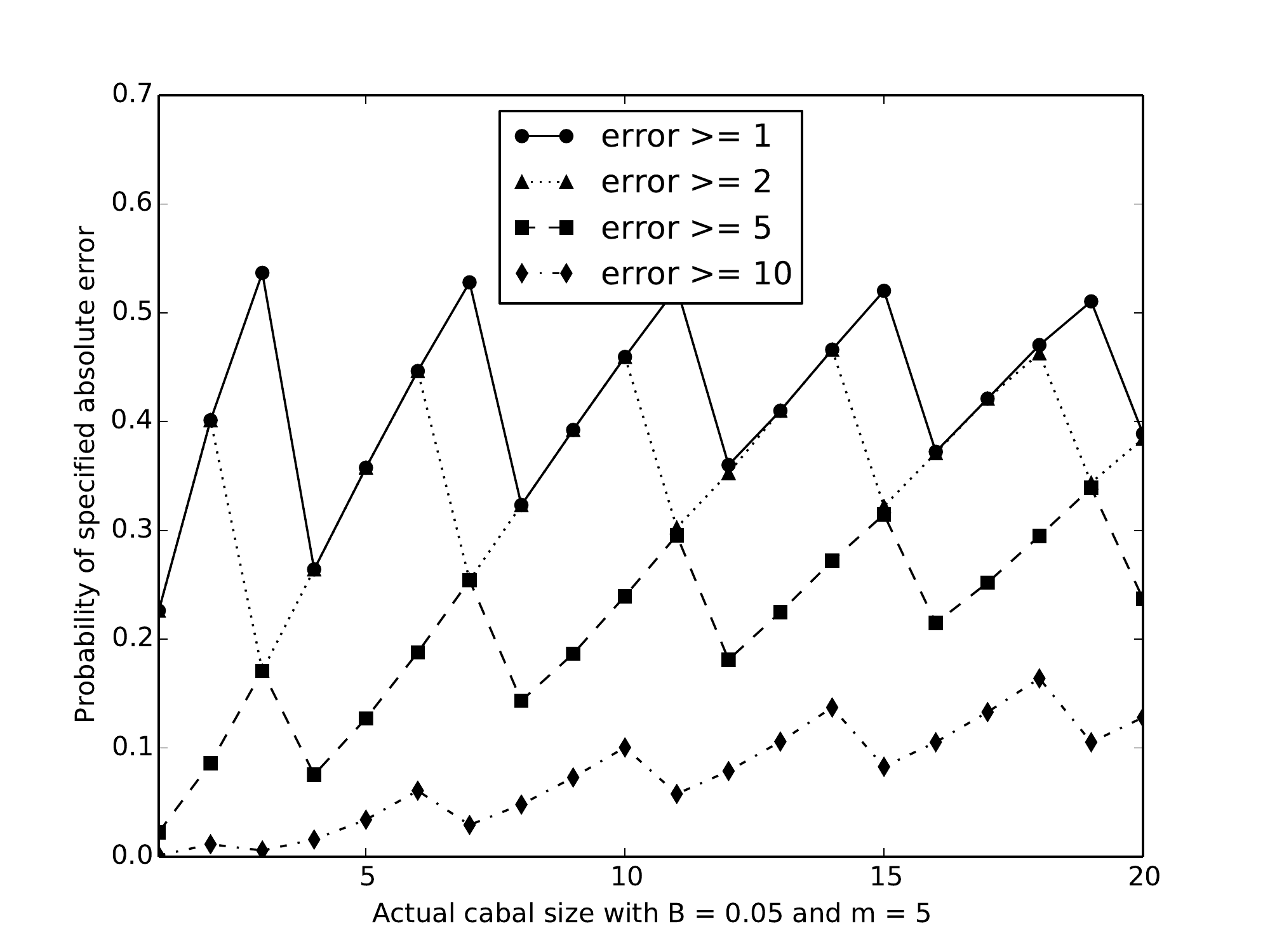}\includegraphics[height=0.13\textheight]{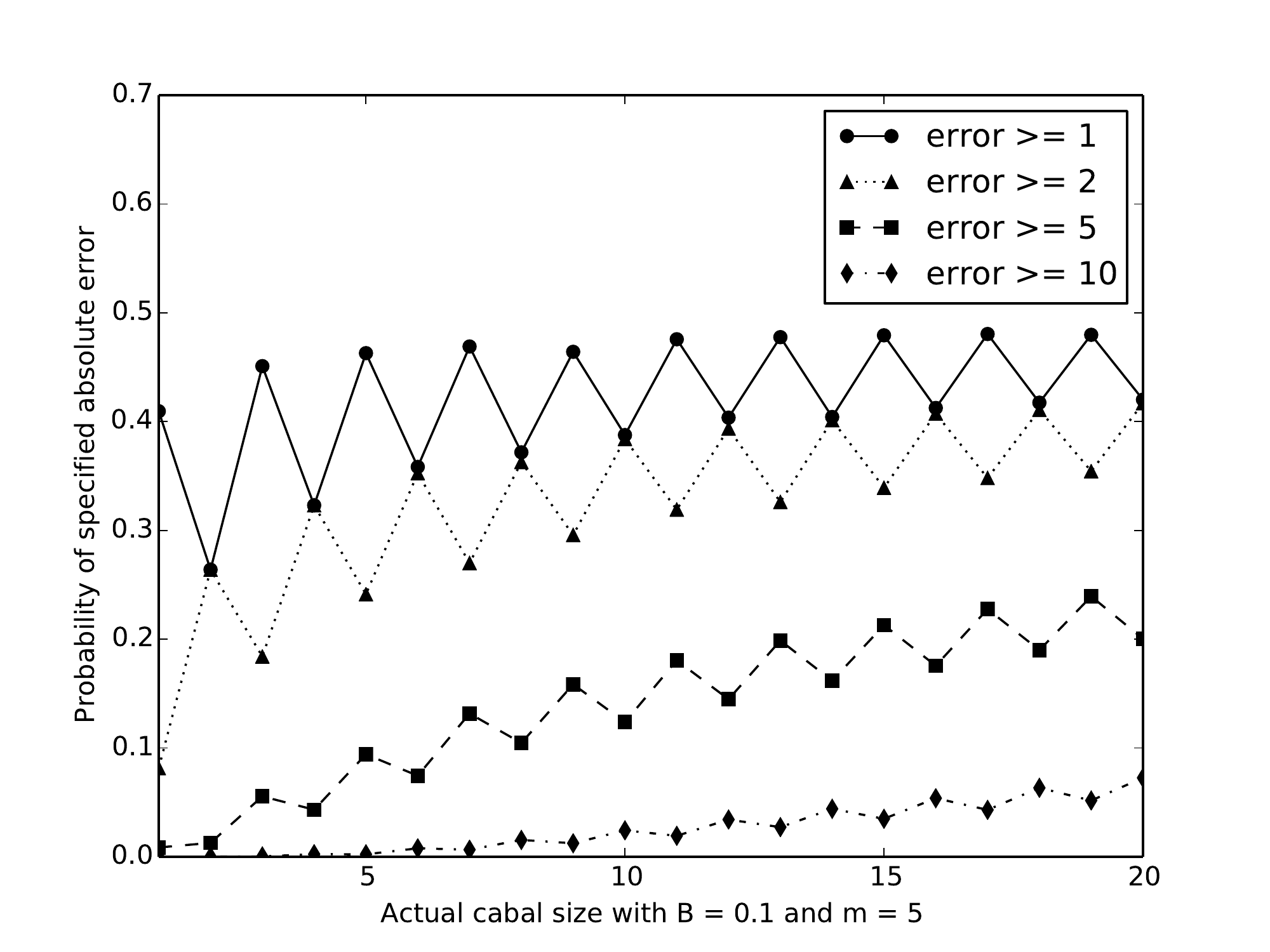}\includegraphics[height=0.13\textheight]{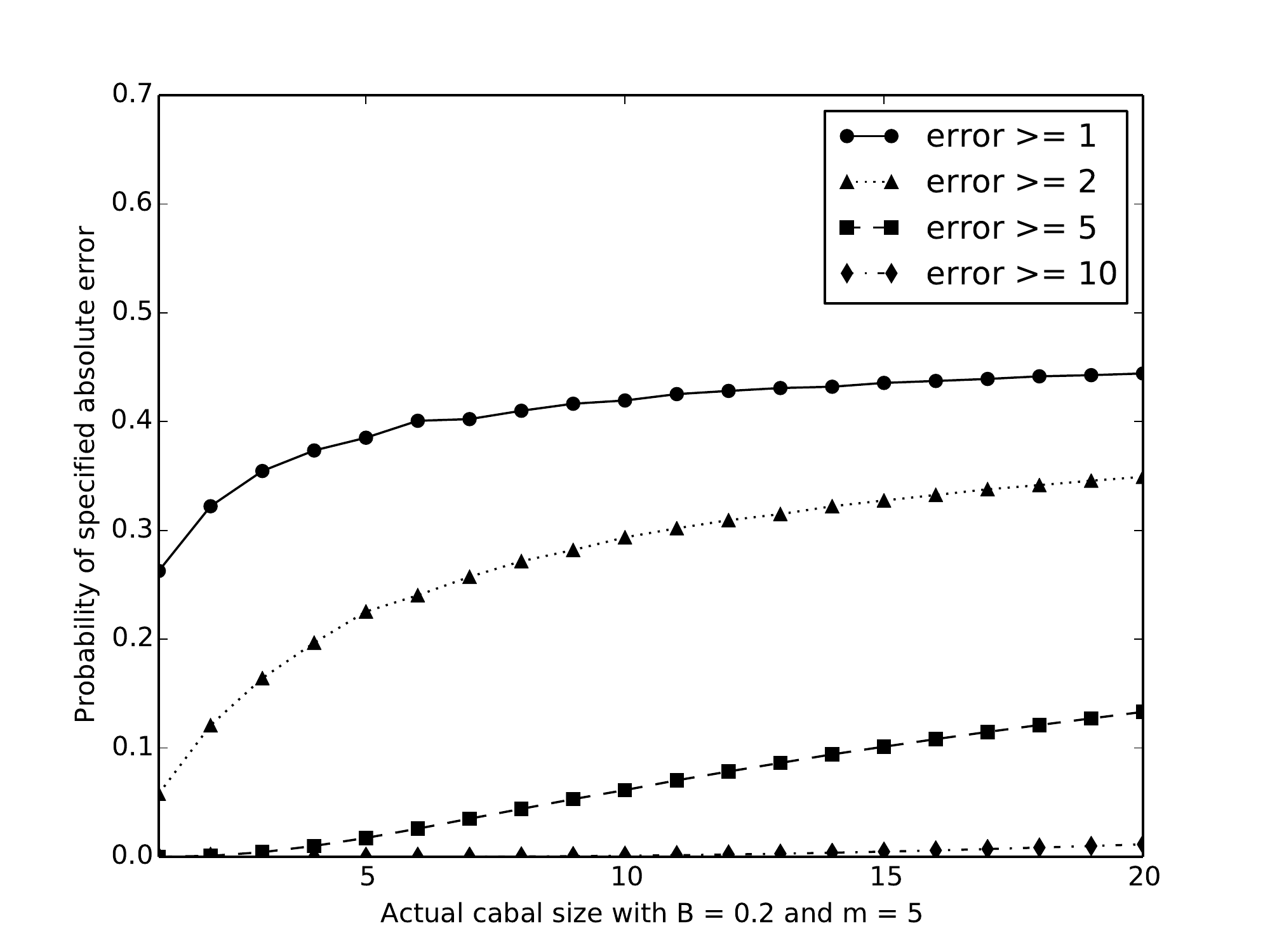}\includegraphics[height=0.135\textheight]{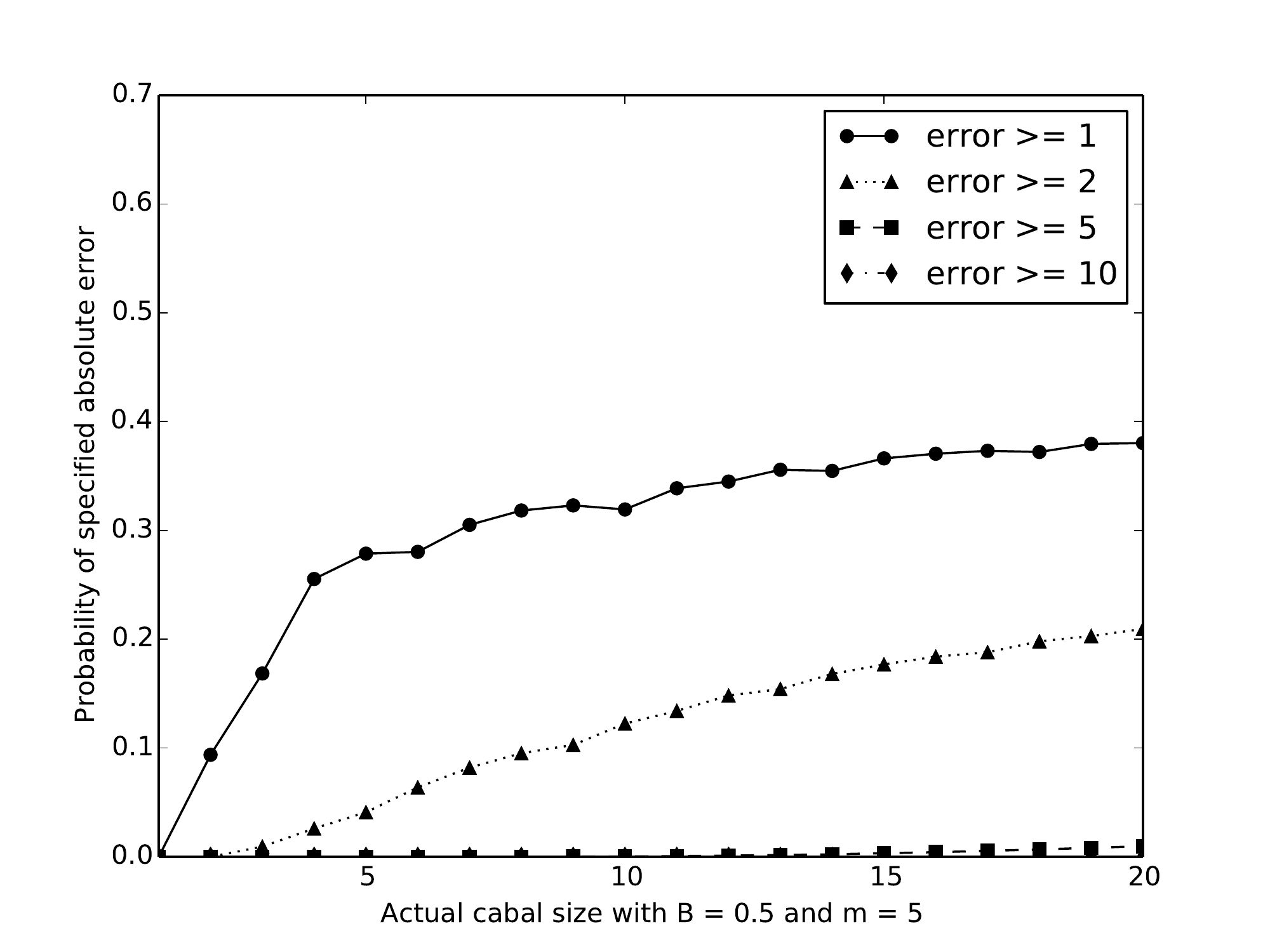}
\caption{Probability MLE error is at least a specified value ($1$, $2$, $5$, and $10$; different curves in each subplot) as a function of true cabal size $c$.  Number of meetings $m$ is $1$, $2$, or $5$ (top to bottom); the fraction $B$ of middle-relay bandwidth controlled by the adversary is $0.05$, $0.1$, $0.2$, and $0.5$ (left to right).}\label{fig:dmle-abs-error-prob}
\end{centering}
\end{figure*}

From these distributions we may compute, for each parameter vector, the probability that the adversary will compute an MLE of cabal size that has error at least a certain size.  Figure~\ref{fig:dmle-abs-error-prob} presents a matrix of subplots that show these probabilities.  In each subplot we show, as a function of the true cabal size $c$, the probability that the adversary's MLE computation differs from $c$ by at least $1$, $5$, and $10$ (different curves within the subplot).  This is repeated across the matrix for different values of $m$ ($1$, $2$, and $5$, left to right) and $B$ ($0.05$, $0.1$, $0.15$, $0.2$, $0.5$, $0.75$, and $0.9$, top to bottom).

From Figs.~\ref{fig:basic-emle}--\ref{fig:dmle-abs-error-prob}, we see that MLE is not an unbiased estimator of cabal size but that, in expectation, it gives a fairly close estimate.  While the distributions may be spread out and discontinuous, by the time the cabal has $m=5$ meetings and the adversary controls $B=0.2$ of the middle-relay bandwidth, the MLE distribution is starting to converge and the probability of a substantial error is fairly small.

%% file: profile.tex
\section{Example: Picking Ripe Onions}\label{sec:profile}

The set of users of a particular site may be similar to a cabal
communicating via multicast or IRC\@. While they may not be holding
simultaneous meetings or even see themselves as a group, an adversary
may target them because they are users of that site.  A site may be interesting for exogenous reasons, \eg, because of a public mention of it or its presence in the browser bookmarks of a previously compromised target.  The adversary may want to count the users visiting a site that he has targeted, but site popularity may also be a criterion for targeting a site.  For example, he might target onionsites that show a sudden spike in popularity.

Our analysis here essentially applies to Tor users visiting many ordinary Internet sites, but we focus on onionsites, particularly hidden web services.  These were
designed to hide many features typically visible for ordinary
websites. They have also had recent design changes specifically
intended to make it harder for an adversary to discover a site's .onion
address, popularity~\cite{prop224} or network
location~\cite{prop247}.

Beyond this inherent interest, such sites are plausible candidates for
targeting of their users.  Onionsites offer protections for their users
that are not yet common on the Internet, much like HTTPS-protected
websites when those were relatively rare.  Like HTTPS-protected
sites of that era, there has been a prominent perception that most current
(thus still early-adoption) onionsites are intentionally set up to
thwart targeted attack.  Even if that perception of
onionspace is initially inaccurate, the perception can itself
drive attention of attackers, thus becoming somewhat self-fulfilling.

As noted, learning about site popularity may be an adversary goal for
a targeted site or may be a criterion for deciding to target a site.
Previous work~\cite{Biryukov-2013,owen-2015} has measured popularity
of onionsites by requests to the onion address directory.  Besides
the directory system design changes that make this approach much less
feasible, it can also be a
misleading and inaccurate measure of onionsite
popularity in several ways~\cite{thoughts-on-hs}.
We can, however, use variants of
the techniques described in previous sections to measure onionsite
popularity (as well as the popularity of other Tor-accessed Internet sites).

Directory requests and even site connections can be unreliable
indicators of human interest in onionsites because of crawlers and
automation such as link prefetching.  Thus, just knowing the number of connections or distinct clients connecting to a site is not very useful to a targeting adversary.  Information about the distribution of
connections among users would be much more useful both for
understanding a site's popularity and for selecting its users for
targeting.

Unlike the cabal-meeting case, because visits are not synchronized
the targeting-adversary
technique of starting with middle relays might seem problematic for
individuating onionsite visitors. But, as with our analyses above,
distinct guards are a fairly accurate indicator of distinct clients up
to a moderate-size set of clients, all the more when each client
uses a single guard per destination (or overall).

Even if site users do share guards, as long as guard overlap is
infrequent enough, this will still give a targeting adversary
a much better idea about interest in the site than could be
obtained via previously published techniques. ``Infrequent enough''
implies that the rough picture of popularity painted by targeted-site
connections per guard per unit time presents a ballpark estimate of
site-user activity distribution.  

To use our middle-relay techniques, we must assume that destinations
of potential interest can be recognized by an adversary in the middle
of circuits. We do so based on traffic patterns plus possibly other
factors like latency, which has been known since at least 2007~\cite{ccs07-hopper} to leak some
identifying information for Tor communication. Destination fingerprinting of
route-protected communications predates Tor~\cite{hintz:pet2002}. How
effective it is has been the subject of much
research~\cite{wang-finger:popets2016}.

Though the level of success for fingerprinting Tor destinations in
general may remain uncertain, onionsites constitute a much smaller
set. And current onionsite protocols are sufficiently different from
applications connected over vanilla Tor circuit protocols
that separating these is currently very easy and reliable
for relays carrying their streams. There are about 60,000 unique onion
addresses in the directory system, but of these typically only a few
thousand at any time are reachable and return a connection to a web
server. If these numbers are roughly representative (even if we add
in onionsites not listed in the directory system and those that can
only be reached if proper authentication is shown to the Introduction
Point), then fingerprintability of all persistent onionsites becomes a reasonable assumption. Note that our techniques are not
affected by whether or not a website is listed in the onionsite
directory system or requires authentication to be reachable.  The most direct technique for a middle relay targeting adversary is
then to simply count the number of connections going to a particular
onionsite from each guard. This will already give a rough picture
of the distribution of client activity as well as which guards are
most worth targeting for further adversary interest.

\subsection{Recounting Onions From Our Past}

Given our assumptions about numbers and fingerprintability of .onion
websites, and rough numbers and distribution of their users and
implications for individuating clients by guards, there are additional
estimation techniques at our disposal.

We can estimate the number of clients visiting a site $n$ or more
times using capture-recapture techniques. These were originally used
for species population estimates in biology where it would be
impossible to observe the entire population, e.g., estimating the
number of fish in a given lake. They are now used in many settings,
including computer networks~\cite{capture-recapture-in-networks}.

The Lincoln--Petersen (LP) estimate is the simplest and quickest for
our adversary to perform, provided we add a few assumptions to those made above. To calculate the number of clients making
$n$ or more connections to a target site per unit time, we must
assume (1) that all clients (targeted or
otherwise) visit a target site with the same frequency during different
sampling intervals of the same length, and (2) that connections
(visits) and observation intervals are such that no connection is
counted in more than one interval.  To avoid problems with division by 0, we use the Chapman version of the LP estimator, given by $\hat{N} = \frac{(c_1 + 1) (c_2 + 1)}{m_2 + 1} - 1$, where
$c_1$ is the number of guards in interval $t_1$ to be `captured' and
`marked' as carrying circuits for $n$ or more visits in that
interval, $c_2$ is the total number of guards observed to `visit the
target' $n$ or more times in $t_2$, and $m_2$ is the number of guards
`marked' in $t_1$ and observed visiting the target at least $n$ times in $t_2$.

Note that, if multiple clients used the same guard simultaneously to connect to an onionsite through one or more compromised middles, the adversary might reasonably conclude that the guard is in fact serving multiple visitors to the onionsite.  This does not follow with certainty, and we do not model such reasoning here.

\begin{figure}
\includegraphics[width=0.5\textwidth]{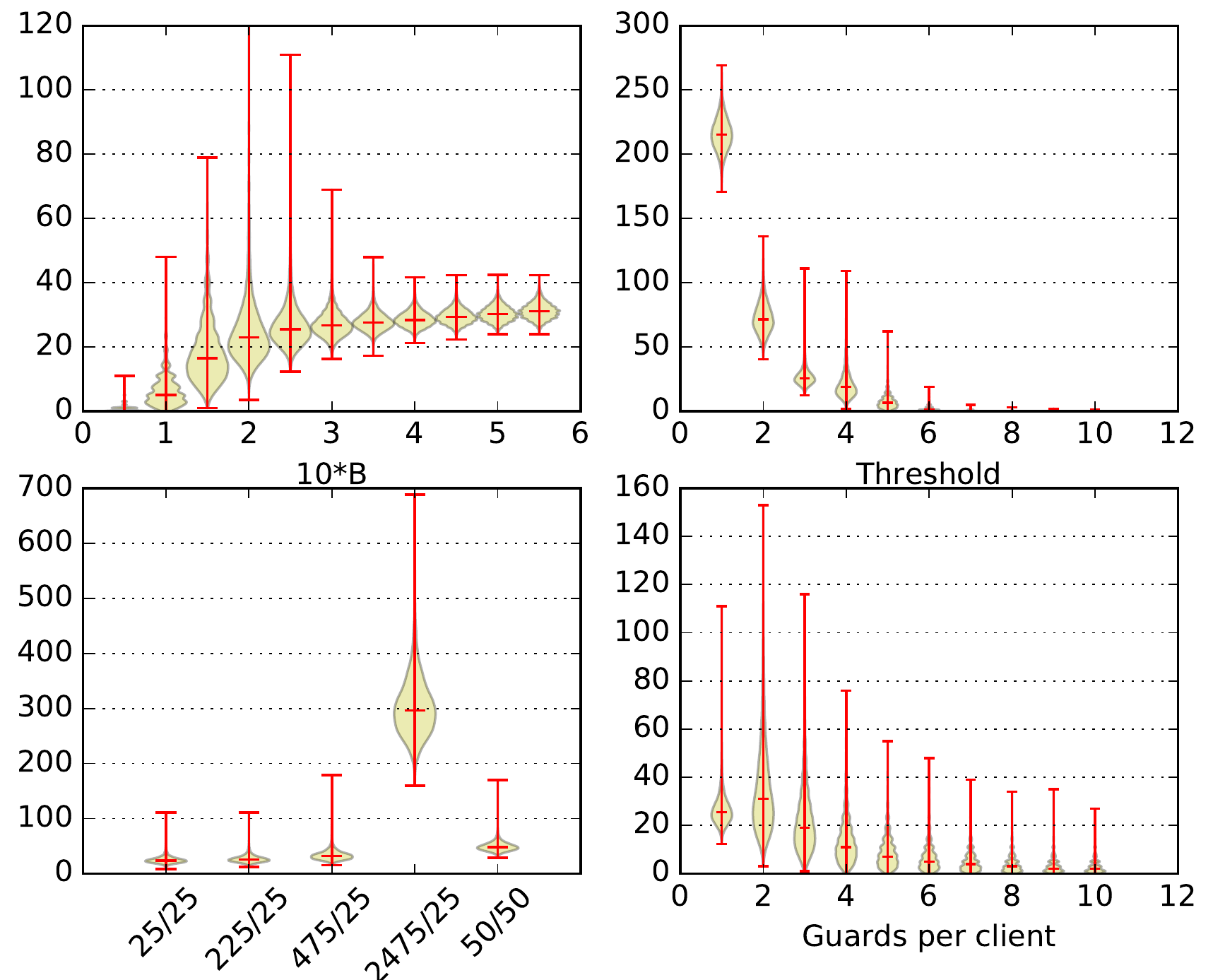}
\caption{Violin plots showing Chapman estimates for 10,000 trials each of: different values of $B$ (top left), different threshold values (top right), different combinations of regular/targeted client types (bottom left), and different numbers of guards per client (bottom right).  When not being explicitly varied as noted, the parameters are: $B=0.25$, threshold of 3, 25 targeted clients who visit 10 times per period, 225 regular clients who visit 2 times per period, and 1 guard per client.}\label{fig:lp}
\end{figure}

In Fig.~\ref{fig:lp} we show the results of numerical experiments with the Chapman estimator.  In these experiments, we assume 2,500 guards and 5,000 middle relays; each of the latter is compromised with probability $B$.  We run each experiment 10,000 times.  For each client, we pick a guard set.  We assume there are two types of clients: ``regular'' clients who visit the targeted site twice during each examination window and ``interesting'' clients who visit the targeted site 10 times during each examination window.  Each experimental run specifies the number of each type of client; for each client, we have it repeatedly pick (twice or 10 times, depending on its type) a guard from its guard set and a middle relay.  We track the number of times that each guard is used with any compromised relay; this could be multiple compromised relays, and the same guard could be used by multiple clients.  We specify a threshold value; guards that are seen by compromised middle relays at least this many times are considered marked and are remembered by the adversary.  This process is repeated again to model the second examination window.

The subplots of Fig.~\ref{fig:lp} show violin plots that focus on varying different parameters; except for the parameter being varied in a particular subplot, these use $B=0.25$, a threshold of 3, a client mix of 25 targeted clients (who each visit the destination 10 times per period) and 225 ``regular'' clients (who each visit the destination 2 times per period), and 1 guard per client.  The subplots examine the effects of varying $B$ (top left; results for $B = 0.05$, $0.1$, $0.15$, $0.2$, $0.25$, $0.3$, $0.35$, $0.4$, $0.45$, $0.5$, and $0.55$), the threshold (top right; results for thresholds of 1, 2, 3, 5, 7, and 10), the client mix (bottom left; results for regular/targeted clients combinations of 25/25, 225/25, 475/25, 2475/25, and 50/50), and the number of guards per client (bottom right; results for 1, 2, 3, 5, and 10 guards per client).

Considering Fig.~\ref{fig:lp}, we see that the adversary's estimate increases in accuracy with $B$, as expected, but much of the gain comes in ensuring that the adversary controls one fifth to one third of the middle-relay bandwidth.  We also see that the best threshold seems to be 3 for this combination of parameters; this also appears to be the case for other parameter combinations that we explored (with the regular clients visiting the destination twice and the targeted clients visiting 10 times).  As might be expected, increasing the number of guards generally decreases the adversary's accuracy.  Finally, the adversary's estimate is reasonable for small enough numbers of targeted users; however, when these users are just 1\% of the total user population, the estimate is off by a large amount.

More sophisticated and accurate estimators exist. For example, the
Schnabel estimator generalizes Lincoln--Petersen to incorporate
repeated samplings with marking of not-previously-marked members of
the population. Other estimators may allow relaxation of assumptions
such as constant ratios from one sample to the next.  And, unlike most
applications, the set of unmarked guards has a structure that can
inform our estimates.  For example, given a guard observed to have
carried circuits connecting to a targeted site 9 times in a day, we
can use the expectation that the associated client actually visited 10
or more times that day to adjust our estimates, and similarly for
smaller numbers of target visits per guard per day.

%% file: discussion.tex
\section{Discussion}
\label{sec:discussion}

In this paper, we have considered adversaries with different goals and
strategies, but often with the same endowment and capabilities as
adversaries in previous work. Another important difference only touched
on above is how long an
adversary may have (or be willing to use) some of his resources.
This can affect both attack success and decisions about which attacks to
attempt.

We now briefly describe some of the temporally dependent
features of an adversary's endowment and strategy, although we leave detailed analysis of this for future work.
We then describe possible countermeasures to a
targeting adversary, particularly one with temporal limitations on his
endowment.

\subsection{Time Is on My Side}
\label{subsec:time}

Temporal aspects of adversary endowment have been considered for a
long time~\cite{ostrovsky-yung-91}, and have also been applied to
onion routing before Tor~\cite{strl01correlation}. And we have
noted that intersection attacks are inherently across
time. Nonetheless, these adversaries have a uniform bound on
resources; what varies is either the distribution of these~\cite{ostrovsky-yung-91} or of network resources and
status~\cite{jwjss13ccs}.  And generally, these adversaries
will still be equally happy to attack any user, or possibly any user
with the same behavior on the network.

In this section we discuss a Tor adversary that has, or is willing to
deploy, different amounts and types of resources at different times,
typically based on some particular target.  Limitations on endowment
could derive from funding or statutory start date or expiration. (We
assume that the adversary has enough understanding of his funding
status to plan.)  More generally, given a level of required
resources, approving or committing to any course of action is
typically easier if the anticipated duration of needed resources is
relatively clear, whether for oneself or when authorizing others, and
whether psychologically or in policy-based decisions; the shorter
that duration is, typically the greater the willingness to commit
resources. Similarly, the presence of intermediate goals, which if
achieved would themselves be of interest or would help
determine whether and how to commit to further action, can make
commitment to an action more acceptable hence more likely.

Johnson \etal~\cite{jwjss13ccs} set out as a behavioral user class an
IRC user, who 27 times a day creates the same single IRC sessions to
the same server. We now compare their analysis of IRC users' security
against an end-to-end correlating relay adversary to the adversary in
Sec.~\ref{sec:irc} targeting a cabal that meets on a private IRC
channel.  For this user type, they considered variously endowed
adversaries and looked in
detail at a relay adversary allocated 100 MiB/s, approximately $4\%$
of the total network bandwidth at the time of their analysis, and on
the order of the largest identified families of relays under a single
operator or organization. The time until an IRC user experienced the first
end-to-end correlation failure (both guard and exit of a circuit
compromised) was analyzed under an optimal distribution of adversary
bandwidth to guards and exits (roughly four to one). 
In our scenario, a relay is assumed to be able to always identify a
cabal connection passing through it. So we should assume that the
Johnson~\etal\ adversary is able to devote all relays to the guard
position. We thus very roughly estimate a 20\%
reduction in median time to compromise compared to what Johnson~\etal\
reported.

For a cabal size of 10 or 20, and a roughly comparable fraction of
bandwdith allocated to middle relays, our targeting adversary will
have a good idea of cabal size and will identify the guards of
nearly all cabal members in under 4 days (100 meetings). According
to the analysis by Johnson et al., the just-described
contributed-guard adversary will require about 10 times as long to get
a much rougher idea of cabal size, having identified guards for
roughly half the cabal. To get approximately the same likelihood as
the targeting adversary of having identified guards for almost all of
the cabal will take 40-50 times as long (under a week vs.\ 150-200
days).  This is not just about size: in 
about a week of IRC usage the
targeting adversary will have a good sense of cabal size, cabal
guards, and client send-receive activity per cabal guard (which may
indicate cabal leaders and \emph{will} indicate which members send the
most). 

The contributed-guard adversary is also no more likely at any time to
identify a cabal leader than any other member. This also holds for
the targeting adversary with respect to identifying a leader's guard.
But it will still typically take less than a week at the stated rate
of meetings.  For a leader recognized by message patterns, a
determined targeting adversary can bring all resources and attacks to
bear to significantly increase his chance of bridging a leader's
guard.  Some attacks may take weeks to know if they have
succeeded, but others will take only hours or even minutes. A leader's
guard might be resistant to all attempts at compromise or
bypass. But such relays can then be subject to persistent DoS or other
resource depletion attacks~\cite{sniper14}, giving the adversary
another chance with a new leader guard.

One guard vs.\ three or more guards in general involves many
trade-offs, some of which we have identified. Our analysis of leader
identification in Sec.~\ref{sec:irc} clearly favors using single
guards to thwart our targeting attacker. But vs.\ the
contributed-guard adversary (which was a primary basis for Tor's
decision to move to single guards~\cite{one-fast-guard}), this is
much more significant in terms of speed of attack and adversary
feedback in process---despite a correlating adversary
automatically identifying any client for which he owns the
guard. Though the correlating adversary remains important and should
not simply be abandoned, this again shows that even when they agree
about what is more secure, the targeting adversary is often the
dominant basis for the assessment.

Johnson et al.\ also assume a network at steady state, after adversary
relays have obtained the guard flag. Middle relays can see some usage
in less than a day after announcing themselves and reach steady state
in about a week. Relays generally take over a week to obtain the guard
flag and about ten weeks to reach steady state for guard
usage~\cite{torblog-relay-lifecycle}. For an adversary 
mounting an attack from scratch, the above time comparison thus
overstates significantly in favor of the contributed-guard adversary.

\subsection{Possible countermeasures}
\label{countermeasures}

The primary purpose of this paper is to describe a class
of adversary that has been overlooked but we believe is as
pertinent or more pertinent for Tor than the one generally receiving
the most attention.  Before concluding, however, we wanted
to at least sketch some possible ways to improve resilience against
such selective adversaries on the Tor network. In the interest of
space (and time) we have limited the scope of our analysis to
adversaries at Tor relays, only minimally considering an adversary on
the network links between them and/or between clients or destinations
and the Tor network. Any countermeasure we describe
here will likely need to be significantly redesigned to be effective once
those resources are added to the adversary arsenal. So there is little
point to providing more than a sketch here.

\noindent{\bf Layered Guards:} The same paper that introduced Tor guards also
introduced the idea of layered guards~\cite{hs-attack06}. The notion
of layered guards has been revisited periodically, most recently in a
Tor Proposal on slowing guard discovery for onion
services~\cite{prop247}. Using only one or a small set of relays
for each client-guard's middle could make it hard to
identify guards just as guards make it hard to identify clients. But,
if single persistent middles are chosen, then randomly selected exits
could possibly enumerate and monitor the behavior of associated users
for many destinations and, worse, will always know for which
destination without 
fingerprinting. In the case of
fingerprinted onion services, cabal/user enumeration will be possible
using (easily spun up) middles chosen as rendezvous points.  In
general, the number and rotation of guards and their second-layer
guards can complicate determination of cabal size as well as guard discovery.

\noindent{\bf Randomized selection of guard set size and duration:}
As our analysis shows, single, persistent guards generally provide
much more enumeration information about size of a cabal or set of
targeted-site users and
more information about targeted-site user behavior than does a set of
guards with more overlap between clients' sets. And long
persistence means that bridging an identifed guard is not needed for
monitoring a portion of targeted client's behavior, and that, if a
bridging is attempted, it will
pay off for a long period of monitoring targeted-client behavior and
IP address(es). And that bridging need not be quickly successful
to be useful. In addition to enlarging and randomizing size
of a client's guard set, selecting
guards for less persistent or predictable periods would also counter
targeting attacks, pseudonymous profiling, and confidence in the expected
value of bridging a guard. Other related strategies may be
worth investigating, such as a client limiting and controlling the use
of the same guard for visits to the same sensitive site.
Obviously there is a tension between the increased risk of correlation
attack from using more guards for shorter periods and targeted attacks
on cabals or clients of targeted sites.

\noindent{\bf Trust:} One way to simultaneously reduce vulnerability from both
targeted middle-relay attacks and untargeted correlation attacks is to
incorporate trust into route selection. The paper that introduced
guards for Tor observed that guards could be ``chosen at random or
chosen based on trust''~\cite{hs-attack06}. Subsequent work noted that
trust could be based on many of the criteria just mentioned as useful
for determining which bridging strategies might be effective against
which guards and introduced a mathematical characterization of trust
as the complement to the probability of
compromise~\cite{trusted-set}. The downhill algorithm~\cite{jsdm11ccs}
explored combining layered guards with trust: A the first relay in a
circuit is chosen from a small set of those most highly
trusted. Later relays in a circuit are selected from ever larger sets
that include relays further ``down the hill'' of assigned trust.
This would add delay to enumeration attacks as well as reducing their
effectiveness. It could also be combined with varying periods of guard
rotation in various ways. An unpublished version of the downhill
algorithm had a slowly rotating first-hop relay set and ever faster
rotating relay sets for each subsequent hop~\cite{PC}. 
For trust-based protections to be effective in practice, trust
of all elements in the network path (ASes, IXPs, submarine cables,
etc.) must be considered~\cite{jjcsf15popets}.

\noindent{\bf Standardized onion service traffic templates:}
Our targeting attacks on onion services are dependent on
the effectiveness of fingerprint-based individuation of them.  Making
traffic fingerprints of many onion services similar to each other
could reduce the effectivness of those attacks.
Providing simple bundles or templates for users wanting to set up
onionsites would be useful for many reasons, incorporating data
management and communication protocols to create default standardized
traffic fingerprints among them. To reduce the likelihood that a
single draconian standard will discourage site operators from using these
defaults, a small number of templates might be available depending
on basic site configuration choices. Sites using standardized templates
can also communicate this to clients either in their directory information
or upon first contact. Traffic fingerpint normalization can then be
enhanced by cooperation between clients and sites. To further
facilitate adaptation to individual sites, onion protocols could
use componentized chunks of fingerprint-normalized communication,
possibly split over multiple circuits. Again the trade-offs against
correlation vulnerability would need to be considered, but we hope
we have shown that fingerprinting by interior elements of the network
is as realistic or more realistic and serious a threat to Tor's most
sensitive users.

%% file: conc.tex
\section{Conclusions and Future Work}

We have introduced targeting adversaries, who focus on particular
system users or groups of users, and shown that the difference in
goals and strategy of these overlooked adversaries can lead to attacks
on Tor users at least as devastating and relevant as any Tor attacks
set out in previous research.  While we have shown the capabilities of
a targeting adversary in realistic scenarios involving a published
multicast system, IRC, and onionsites, we cannot hope to quantify in
this introductory treatment the possible effects for all Tor users,
for all possible client locations and network destinations.  We
anticipate extensive future research into targeting adversaries,
including: abstract characterization and formal treatment of targeting
adversaries, expansion of contexts in applied targeting adversary
models (such as network link adversaries), and analysis of security
against combinations of targeted and hoovering adversaries,
particularly when large and well-resourced. And we expect all of
this to have an impact on future guard and path selection algorithms.

Onion services are also an active area of discussion and
redevelopment~\cite{prop224,prop260}.  We expect that an interesting
and useful direction for future research will be the analysis of the
effects of different redesign proposals on security in the
context of targeting adversaries.  This will require a substantial
extension to TorPS~\cite{torpscode}, which does not currently support
modeling of onion services.

%% file: model.tex
\section{Model}\label{app:model}

The model we present is meant to fit within general models of network
actors. Jaggard~\etal~\cite{jjcsf15popets} describe a more general
model and associated formal language. We focus here in setting out
a model specific to Tor and that is suitable for describing targeting adversaries. Adversaries are
modeled in App.~\ref{app:adv}.

\subsection{Entities}

\begin{description}
    \item[Principals] \princ\ is the set of all principals.  We think of these as the individual humans using the system.
    \begin{itemize}
        \item These may be mobile or using multiple Tor client instances
        \begin{itemize}
            \item For now, we assume that a change in location by a principal implies a change in the client instance being used.
        \end{itemize}
        \item The principals are the ultimate targets of attacks.  A natural simplifying step is to focus on attacking Tor clients instead of the principals using the clients.  Doing so omits the consideration of attacking a principal who uses Tor in multiple locations because, in our current model, each location would have a different client.
        \item Principals might use multiple identities, \eg, via the \textsf{New Identity} button in the Tor Browser
    \end{itemize}
  \item[Identities] \identities\ is the set of all identities. An
    identity is any collection of actions or properties that might be
    tied together under a pseudonym from some perspective. But unlike
    the usual usage of `pseudonym', we consider the possibility that a
    single identity might include the behavior of multiple
    principals. In practice, we generally make the simplifying
    assumption that relevant perspectives assign one principal per
    identity. Clicking the \textsf{New Identity} button in Tor Browser
    results in a new identity from the perspective of a destination
    seeing prior and posterior connections from that client.
    \item[Clients]  \clients\ is the set of Tor clients.
    \begin{itemize}
        \item To simplify things, we assume that each client is used by a single principal.  As noted in the discussion of principals, a single principal may use multiple Tor clients.
    \end{itemize}
    \item[Relays] \relays\ denotes the set of all Tor relays.
    \item[Guards] \guards\ denotes the set of all Tor guards.
      $\guards \subseteq \relays$. We use \guardset\ for a set of
      guards used by an entity. We will assume that each client
      chooses a single set \guardset\ of guards from \guards\
      according to a distribution \gdist, which may be time
      dependent. More complex guard selection strategies are also
      possible. \cite{guardsets-pets15}
  \item[Middles] \middles\ denotes the set of all Tor middle
    relays. $\middles \subseteq \relays$
  \item[Exits] \exits\ is the set of Tor exit nodes.  $\exits
    \subseteq \relays$.
  \item[Circuits] \circs\ is the set of Tor circuits.
  \item[Destinations] \dests\ is the set of destinations.
  \item[Links] \links\ is the set of network links between relays,
    bridges, clients, and destinations. We will consider only links
    between clients and the Tor network or destinations and the Tor
    network in this paper, and only abstractly. In practice it can be
    further decomposed into ASes, ISPs, submarine cables,
    etc. \cite{jjcsf15popets}
  \item[Bridges] We will not analyze the use of Tor bridges in this
    paper, and include this entry only for readers who may be
    wondering about the omission.
\end{description}

\begin{remark}[Conventions]
We typically use lowercase, italicized letters for atomic individuals (\eg, a guard $g$), uppercase, italicized letters for distinguished sets of atomic individuals (\eg, a principal's guard set \guardset), and uppercase, caligraphic letters for the set of all such possibilities (\eg, the set \guards\ of all guards).
\end{remark}

\subsection{Dynamics, history, and activity}

Roughly, we expect a principal to choose a client \cl\ (possibly multiple clients over time).  Each client will have a distribution \ddist\ on destinations (this may be inherited from the underlying principal).  Given \ddist, the client will choose a distribution \gdist\ on $2^\guards$; this may be time dependent.  (We expect that \gdist\ will capture the various guard-selection procedures of interest.  Its support may thus be limited to all sets of size $k$, \etc.)  The guard set may change over time; indeed, the effect of temporal change is a question upon which this work will shed light).

Once the client has chosen her guard set, she selects/receives destinations one at a time.  We assume that the destination $d$ is drawn from \dests\ according to a distribution \ddist\ that depends on either the principal or the client, depending on which type is the ultimate target of attack.\footnote{\ddist\ could be enriched to model the fact that destinations might not be chosen independently of each other.  One destination may induce traffic to/from another destination automatically, or the traffic may prompt the user to explore other destinations on her own, \eg, going to the Wikipedia page for a person mentioned in a news article that she reads.  \ddist\ might then produce a set of destinations, each of which is fed into the rest of the process modeled here.}  For each destination, we assume that the client selects an existing circuit or builds a new one; we assume that the circuit-selection and -construction decisions are made in an online way, before the next destination is selected/received.  Once the circuit (which we denote \ci) is chosen/constructed, the client and the destination exchange traffic over the circuit.  This traffic may depend on the circuit and destination $d$, and we assume it is timestamped.  As needed, we denote this traffic by $\tr(\cl, \ci, d, t)$.

We capture this flow as shown in Fig.~\ref{fig:modelflow}, with right arrows denoting the sequence of events and left arrows indicating randomized choices.
\begin{figure*}
\begin{centering}
\begin{equation}
p \longrightarrow \cl\ \longrightarrow \ddist\ \longrightarrow\ \gdist\ \longrightarrow (\guardset\ \stackrel{\gamma}{\longleftarrow} 2^\guards) \longrightarrow (d \stackrel{\delta}{\longleftarrow} \dests) \longrightarrow \ci\ \longrightarrow \tr(\cl, \ci, d, t)
\end{equation}
\end{centering}
\caption{Activity flow in our model.}\label{fig:modelflow}
\end{figure*}

We think of the history and current activity of the principal/client as the primary things that an adversary might use in an attack.  As an initial step, we take the history and current activity to comprise instances of:
\begin{itemize}
    \item Circuit construction, including timestamps (and perhaps some portion of the circuit)
    \item Tor traffic, including some projection of the information in $\tr(\cl, \ci, d, t)$ (the projection may be different for different instances of traffic)
\end{itemize}

%% file: adv.tex
\section{Adversaries and their Goals}\label{app:adv}

\subsection{Adversaries}

Adversaries will vary in the type of resources they control or observe.
They may observe or operate Tor relays, or they may observe all traffic
at an ISP or AS\@. They may observe all traffic arriving at a specific
destination. In addition they may have different endowments of each
type of resource, and they may vary in their goals. We will leave goals
until after setting out the targets to which they may direct those goals.

An adversary will be defined as a tuple of resources of the types
defined in the model, where each entry will reflect the endowment,
either as an absolute number or as a fraction. Elements in the tuple
are not automatically independent or disjoint. For example, one
element may reflect the number of compromised guards and another the
number of compromised middles. How to handle such will depend on the
needs of the analysis being done.

\subsection{Targets and the profiles they generate}

For convenience, we let \targets\ be the set of possible adversary
targets.\footnote{It is natural to let this be the set of all
  principals or the set of all clients, but we do not want to restate
  this inclusiveness at every instance of discussing targets, nor do
  we wish to rule out other collections of targets.}  As we consider
possible attacks on targets, we will think of the adversary as having
``profiles'' of his targets.  We leave the formal definition of these
to specific scenarios; broadly, we think of a profile as comprising
some combination of history and auxiliary information such as
identity, affinity, \etc.\footnote{While we defer the formal
  definition of a profile, we note that it must admit a notion of
  containment so that we can meaningfully write
  $\profile\subseteq\profile'$.}  We consider $\identities \subset
\targets$. This is a proper subset because adversaries may be
targeting a collection of identities and may be specifically targeting
a collection of principals, e.g., all the people who visit a
particular website on a given day or receive email from a particular
identity.

We might consider profiles \emph{generated} by targets, \eg, $\profile
= \profile(\target)$.  While again deferring formal definitions, this
intuitively means that all of the history in \profile\ was generated
by \target, and all of the auxiliary information in \profile\ is about
\target.  Of course, for a collection $\targets'$ of targets, we can
generalize this to saying that a profile \profile\ concerns this
collection, written $\profile = \profile(\targets')$, if
\begin{equation}
\profile \subseteq \bigcup_{\target\in\targets'} \profile(\target).
\end{equation}

Similarly, we will write \act\ for ongoing network activity that the adversary might observe.  Connecting this to specific targets raises the same issues raised by profiles, so we will write $\act = \act(\target)$ to indicate that activity is generated by a particular target \target.

\subsection{Adversary goals}
\label{app:adversary-goals}

We now set out some of the abstract goals a targeting adversary may have.

\begin{description}
    \item[Profile matching]  The adversary has a pseudonymous profile and is presented with new network activity; is this activity part of the profile (or how likely is it to be so)?  \Ie, given $\profile(\target)$ and $\act(\target')$, does $\target = \target'$?
    \item[Profile extension] The adversary has a user profile (not necessarily pseudonymous); can the adversary extend this to include additional information (perhaps of a specified type).  \Ie, given $\profile(\target)$, construct $\profile'(\target) \supseteq \profile(\target)$.
    \item[Knowledge gathering] A client is identified as being of
      interest; what else can the adversary learn about the client?
      \Ie, given \target, construct $\profile(\target)$ (maybe also
      with some auxiliary information $\aux(\target)$ related to the
      target).
    \item[Group identification] Which people (or pseudonymous
      profiles) are behaving similarly, \eg, using the same services?
      (If ``behaving similarly'' is defined by some predicate, how is
      the ease of answering this question related to the specificity
      of the predicate?)  As one example, given a target \target\ and
      a predicate \pred\ (maybe satisfied by \target), describe
      $\left\{\target' \vert \pred(\target')\right\}$ (or a nontrivial
      subset of this set).
    \item[Group enumeration] Whether or not the individual targets are
      known, it may be of interest to determine the exact or
      approximate size of a group target, $\card{\left\{\target' \vert
          \pred(\target')\right\}}$.  This is one of the goals considered in the body of this paper.
\end{description}

An example of a group target is a \emph{cabal} with some properties in
common, e.g., meet together on Tuesdays at noon, all were on the same
high school sports team, etc.

As in earlier research, a primary goal is to associate source and
destination of communication. On the other hand, for a targeting
adversary if the source or destination here is a target, other sources
(respectively destinations) may be ignored as uninteresting. Other
goals also become salient for a targeting adversary. For example,
group enumeration may be useful in itself as a measure of a group
target's significance. Other goals of interest may be similarly about
numbers rather than about source and destination correlation per se:
the fraction of a source target's connections devoted to different
categories of activity, website downloads vs.\ chatting or bandwidth
to route-unprotected sites vs.\ onionsites, \etc

%% file: irccomp.tex
\section{Computations}\label{app:irccomp}

\subsection{Identifying an IRC cabal member}

\subsubsection{One-guard case}

If the cabal leader is using one guard, then with probability $B$ the leader will choose a compromised middle relay during the first meeting, allowing the attacker to learn the leader's guard.  With probability $p_b$, the attacker will learn the leader's address.  Alternatively, the leader does not use a compromised middle relay for the first meeting (which happens with probability $1-B$, or with probability $(1-B)^i$ for the first $i$ meetings) but then uses a compromised middle relay (with probability $B$) for the second (or $(i+1)^\mathrm{st}$) meeting.  Once the compromised middle relay is used, then the attacker learns the leader's address with probability $p_b$.  We note that it only matters when the leader first uses a compromised middle relay---the attacker only has one chance to ``bridge'' the leader's guard; if she fails the first time, then we assume that she is not able to successfully bridge that guard on a later occasion that the leader uses a compromised middle relay.  Thus, we have that the probability of the adversary successfully bridging the cabal leader's single guard is
\begin{multline}
B p_b + (1-B) B p_b 
+ \cdots + (1-B)^{m-1} B p_b\\
= \left[1 - (1 - B)^{m}\right]p_b.\label{eq:one-guard}
\end{multline}

\subsubsection{Three-guard case}

We turn now to the case where each client uses three guards.  Our approach parallels the one-guard case, but the computations becomes more complex.  We describe a general approach that works for any number of guards.  The separates out the adversary's successful identification of the client into different cases, depending on how many guards the adversary unsuccessfully bridges before successfully bridging one of the client's guards.  Together, the probability of success in these cases (up to the total number of guards used by the client) gives the adversary's total probability of success.

In particular, we let $F^n_j(i)$, with $i>0$, be the probability that the adversary makes her $j^\mathrm{th}$ observation of a previously unseen guard $i$ meetings after previously observing a new guard, that the leader is using $n$ guards total (and choosing from these uniformly at random for each meeting), and that the adversary fails to bridge this guard.  This involves the leader doing something other than both choosing a fresh guard (\ie, one the adversary has not previously attempted to bridge) and a compromised middle relay for $i_j - 1$ meetings (each time with probability $1-\frac{n-j+1}{n}B$), then choosing a fresh relay and a compromised middle guard (with probability $\frac{n-j+1}{n}B$), and then the adversary failing to bridge that guard (with probability $1-p_b$).  This gives
\begin{equation}
F^n_j(i) = \left(1 - \frac{n-j+1}{n}B\right)^{i-1}\frac{n-j+1}{n}B(1-p_b).
\end{equation}

We also let $S^n_j(m)$ denote the probability that the adversary successfully bridges the $j^\mathrm{th}$ guard she observes and that she has $m$ meetings in which to do so (after trying and failing to bridge the $(j-1)^\mathrm{st}$ guard.  This involves the leader doing something other than both choosing a fresh guard and a compromised middle relay for $k$ meetings (each time with probability $1 - \frac{n-j+1}{n}B$), $0\leq k\leq m-1$, and then choosing a fresh guard and a compromised middle relay (with probability $\frac{n-j+1}{n}B$) and having the adversary bridge that guard (with probability $p_b$).  This gives
\begin{equation}
S^n_j(m) = \left[1 - \left(1 - \frac{n-j+1}{n}B\right)^m\right]p_b.
\end{equation}

With this notation, we can concisely capture the probability of success for any number of guards $g$ as
\begin{equation}
\sum_{k=1}^g \sum_{0 = i_0 < i_1 < \cdots < i_{k-1} < m} S^g_k(m - i_{k-1}) \prod_{j=1}^{k-1} F^g_j(i_j - i_{j-1}).
\end{equation}
Here, $k$ represents the guard that is successfully bridged, and $i_1$, \ldots, $i_{k-1}$ are the meeting numbers at which the previous guards are identified but unsuccessfully bridged.

In the three-guard case, we obtain the following probability of success for identifying the leader of the cabal:
\begin{multline}
S^3_1(m) + \sum_{1\leq i_1 < m} F^3_1(i_1) S^3_2(m-i_1) +\\
\sum_{1\leq i_1 < i_2 < m} F^3_1(i_1) F^3_2(i_2 - i_1) S^3_3(m - i_2).
\end{multline}

\subsection{Guard collisions}\label{ap:guardcoll}

We show in Tab.~\ref{tab:gcprob} the probabilities of multiple clients sharing a guard for one and three guards per client and various cabal sizes.  This assumes 2,500 guards.

\begin{table}[h]
\begin{centering}
\begin{tabular}{|c|c|c|c|c|c|}
\# Clients$\rightarrow$ & 3 & 5 & 10 & 20 & 25 \\ \hline
1 guard/client & 0.1\% & 0.4\% & 1.8\% & 7.3\% & 11.3\% \\ \hline
3 guards/client & 1.1\% & 3.5\% & 15.0\% & 49.8\% & 66.4\% \\
\end{tabular}
\end{centering}
\caption{Guard-collision probabilities for 2,500 guards, one or three distinct guards per client, and various numbers of clients.}
\label{tab:gcprob}
\end{table}

%% file: dmle.tex
\section{MLE Computations}\label{app:dmle}

Here, we present more detail about the process of computing the MLE of cabal size.

First, we construct the function $L(\vec{x}|\theta)$ capturing the likelihood that the adversary will make a sequence of observations $\vec{x}$ if the actual size of the cabal is $\theta$.  This is $0$ if any of the observations is greater than $\theta$.  Otherwise, it is a matter of choosing, in the $i^\mathrm{th}$ meeting, $x_i$ cabal members to observe, observing each of them with probability $B$, and not observing each of the $\theta - x_i$ cabal members with probability $1-B$.  This is done independently for each meeting, giving us
\begin{equation}
L(\vec{x}\vert\theta) = \prod_{i} \binom{\theta}{x_i}B^{x_i}(1-B)^{\theta - x_i},\ \ x_i \leq \theta.
\end{equation}

We consider actual cabal sizes $c$ ranging from 1 to 20.  For each $c$ and each possible observation vector $\vec{x}$ that could be observed for that cabal size (\ie, iterating over $\{0,\ldots,c\}^m$), we compute the value $\theta(\vec{x})$ that maximizes the likelihood function for $\vec{x}$ by numerically testing values of $\theta$ up to 100.  (If multiple values of $\theta$ maximize $L(\theta|\vec{x})$, we choose the smallest as $\theta(\vec{x})$.)  We also compute the probability, given the actual cabal size $c$, that $\vec{x}$ is observed by the adversary.